\documentclass[12pt]{article}
\usepackage{amsmath, mathtools, color}
\usepackage[utf8]{inputenc}
\usepackage[english]{babel}
\usepackage{natbib}
\setcitestyle{authoryear,open={(},close={)}}

\usepackage{accents}
\usepackage{stackengine}

\usepackage{graphicx, geometry,graphicx}
\usepackage{epstopdf}
        \usepackage{dutchcal}

\newcommand{\be}{\begin{equation}}
\newcommand{\en}{\end{equation}}

\renewcommand{\vec}[1]{\boldsymbol{#1}}
\newcommand{\ii}{\textrm{i}}

\def \curl{\mbox{curl\hskip 1pt}}
\def \Curl{\mbox{Curl\hskip 1pt}}
\def \div{\mbox{div\hskip 1pt}}

\def \tr{\mbox{tr\hskip 1pt}}

\newcommand{\ELd}{{\vec{\dot E}}_{L}}
\newcommand{\DLd}{{\vec{\dot D}}_{L}}
\newcommand{\fa}{\mathcal{A}}

\begin{document}

\title{Wrinkles in soft dielectric plates}

\author{Yipin Su$^{1,2}$, Hannah Conroy Broderick$^1$, \\
Weiqiu Chen$^2$, Michel Destrade$^{1,2}$\\[12pt]
$^1$School of Mathematics, Statistics and Applied Mathematics, \\
NUI Galway, University Road, Galway, Ireland\\[12pt]
$^2$Department of Engineering Mechanics,\\
Zhejiang University, Hangzhou 310027, P.R. China}

\maketitle

%

\begin{abstract}

We show that a smooth giant voltage actuation of soft dielectric plates is not easily obtained in practice. 
In principle one can exploit, through pre-deformation, the snap-through behavior of their loading curve to deliver a large stretch prior to electric breakdown. However, we demonstrate here that even in this favorable scenario, the soft dielectric is likely to first encounter the plate wrinkling phenomenon, as modeled by the onset of small-amplitude  sinusoidal perturbations on its faces. We provide an explicit treatment of this incremental boundary value problem. We also derive closed-form expressions for the two limit cases of very thin membranes (with vanishing thickness) and of thick plates (with thickness comparable to or greater than the wavelength of the perturbation). 
We treat explicitly examples of ideal dielectric free energy functions (where the mechanical part is of the neo-Hookean, Mooney-Rivlin or Gent form) and of dielectrics exhibiting polarization saturation. 
Finally we make the link with the classical results of the Hessian electro-mechanical instability criterion and of Euler buckling for an elastic column.

\end{abstract}


\newpage


\section{Introduction}


When a soft dielectric plate is put under a large voltage applied to its faces, it expands in its plane. 
At first, the expansion increases slowly and almost linearly with the voltage until, typically, a local maximum is reached. 
Then in theory, the voltage drops suddenly, until it reaches a local minimum, rises again, to reach the same level it had at the earlier maximum, and then continues to rise. 
In practice the voltage doesn't drop: it stays at the level of the first maximum while the plate expands rapidly, until it starts increasing again with the stretch.
The membrane is said to experience a \emph{snap-through expansion} \citep{ZhSu10, Rudykh12, DorfOgden14, An2015, Li17}. 
This large and almost instantaneous extension is highly desirable in experiments but is rarely achieved because during the snap-through the elastomer fails due to \emph{electric breakdown} \citep{Blok69, Koh11, Huang12}. 
Graphically, the curve of the electric breakdown crosses the voltage-stretch curve before the snap-through portion is completed.

\begin{figure}[h!]
\centering
\includegraphics[width=0.8\textwidth]{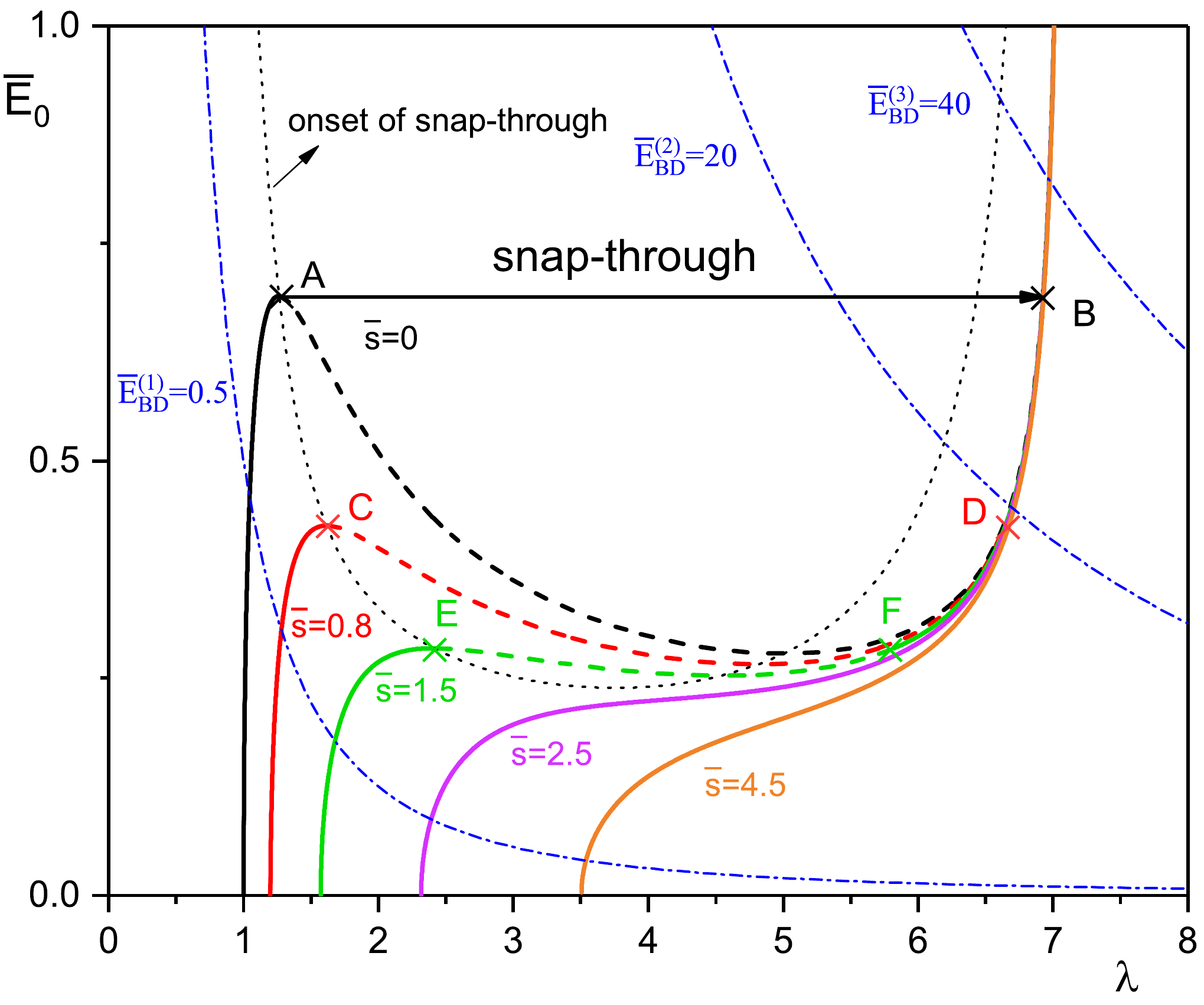}
\caption{
{\footnotesize
Principle of the snap-through giant actuation. Solid lines are the voltage-stretch curves for homogeneous loading at different levels of pre-stress ($\bar s=0,0.8,...,4.5$), when the plate is modeled by the Gent ideal dielectric (here $J_m=97.2$ \citep{Gent96, DorfOgden14}). Dotted line corresponds to the onset of snap-through instability. The dashed parts of the voltage-stretch curves are the theoretical response of the elastomer after the snap-through instability is triggered, which will not happen in practice. The blue dash-dotted lines are hypothetical Electrical Breakdown curves. The situation described by $\overline E_{BD}^{(3)}$ is the most favorable, allowing the initially unstretched material to expand and experience a large snap-through from A to B. For $\overline E_{BD}^{(2)}$, this will not be allowed, but a certain level of pre-stress (here $\bar s=0.8,1.5$) will give a (smaller) snap-through transition (from C to D, from E to F, respectively). 
As the hypothetical $\overline E_{BD}$ curve slides down further, this possibility will vanish eventually (see $\bar s=2.5, 4.5$ curves). For $\overline E_{BD}^{(1)}$, no snap-through is possible.}
}
\label{fig:snap1}
\end{figure}

This undesirable outcome can be avoided in a number of ways, in principle \citep{Koh11, Li11b,Jiang15, Jiang16}. 
We could for instance try to design a dielectric material with a free energy density such that the snap-through sequence is completed prior to electrical breakdown. 
But it seems that such a material has not been synthesised yet. 
We could pre-stretch the membrane so that the snap-through path is shifted below that of the un-stretched membrane. 
But in that scenario the snap-through actuation gain is greatly reduced. 
Moreover, with larger pre-stretch, the corresponding path might become increasing monotonic and the snap-through possibility will then disappear altogether.

These possible events are summarised in Figure \ref{fig:snap1}, where we take $J_m=97.2$ \citep{Gent96, DorfOgden14} as a representative stiffening parameter of elastomers (a different value stretches or shrinks the plots, but the essential results remain the same).
The critical dimensionless electric field of the plate is $\overline{E}_{BD}=V_{BD}/(h\sqrt{\mu/\varepsilon})$, where $V_{BD}$ is the voltage causing electrical breakdown of the elastomer and $h$ is its current thickness. 
This  material constant $\overline{E}_{BD}$ is also known as the  \emph{dielectric strength} \citep{Pelr00}. 
For an equal-biaxially stretched dielectric elastomer, $\overline{E}_{0BD}=\lambda^{-2}\overline{E}_{BD}$, where $\overline E_{0BD}$ is the nominal measure of $\overline E_{BD}$ and $\lambda$ is the in-plane stretch. 
This relation is displayed by the blue dash-doted lines in Figure 1, for hypothetical values of the dielectric strength ($\overline{E}_{BD}=0.5, 20, 40$).

\begin{figure}[ht!]
\includegraphics[width=\textwidth]{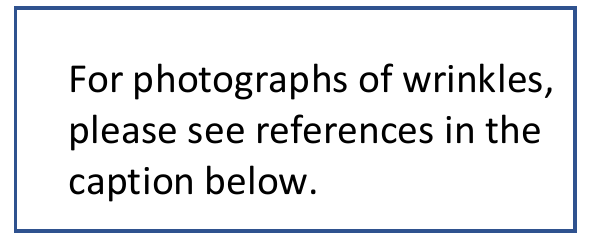}
\caption{
{\footnotesize
Experimental evidence of electro-mechanical wrinkling instability: (a) Collapse of a thin film of the rubber-like material VHB 4905/4910 put under a large voltage \citep{DuPl06};
(b) wrinkling of a VHB 4910 membrane under high voltage \citep{Liu16};
(c) Electric activation of acrylic elastomers \citep{Pelr00}.
We estimate that the ratio of the initial plate thickness to the wrinkle wavelength is $H/\mathcal L \simeq 0.17, 0.35$ in Cases (a) and (b), respectively.
}}
\label{fig-exp}
\end{figure}

In this paper we show that in any case, the snap-through scenario is derailed because the loading curve crosses that of \emph{wrinkle formation} on its way upward.
Indeed, several experiments \citep{DuPl06, Liu16, Jiang15, Jiang16} have shown that  sinusoidal wrinkles appear in soft dielectric plates under high voltage, see examples in Figure \ref{fig-exp}. 
Here we model and predict how they will form.

In Section \ref{section2} we recall the equations governing the large deformation of a dielectric plate subject to pre-stretch and voltage. 

We then rely on the theory of incremental deformations superposed on large actuation \citep{Ogden10, DoOg10, BeGe11, Rudykh11, GeCS12, Rudykh14, Su16, BeSh18} to solve the boundary value problem of small-amplitude sinusoidal wrinkles appearing on the mechanically-free faces of the plate (Section \ref{section3}).
This problem was treated earlier by \cite{Ogden14, Ogden2014} and more recently, by \cite{Yang17} and \cite{Diaz17}, but not in a fully analytical manner as here.
Here we present a general framework to solve the boundary-value problem for a general free energy density. 
We are able to obtain analytical  results in the case of the Gent ideal dielectric, a model which exhibits the typical non-monotonic snap-through loading curve, see Figure \ref{fig:snap1}, and also in the cases of neo-Hookean and Mooney-Rivlin ideal dielectrics.
Thanks to the Stroh formulation and the surface impedance method \citep{Destrade15}, we obtain closed-form expressions for the thin-plate and for the short-wave limits. 
Plotting the two corresponding curves gives a narrow region where all physical plate dimensions and wrinkle wavelengths are located.
We find that it crosses all loading curves before the snap-through can be completed.
We are also able to separate the symmetric and antisymmetric modes of buckling and to solve the dispersion equations in a numerically robust manner.

In the final section (Section \ref{section4}) we present further results, for dielectric exhibiting polarization saturation, and for the specialization of our analytical formulas to known results in classical Euler buckling theory for elastic columns. 
We also make the link with, and extend the Hessian criterion of instability \citep{ZhSu07} for electro-elastic dielectrics.

\section{Large actuation}
\label{section2}


We write the free energy density for the dielectric plate as $\Omega=\Omega({\vec {F}},\vec E_L)$, where $\vec F$ is the deformation gradient and $\vec E_L$ is the Lagrangian form of the electric field $\vec E$: $\vec E_L=\vec F^T\vec E$.
We introduce the following complete set of invariants for an isotropic incompressible dielectric  \citep{Ogden05, Ogden06, Rudykh14, Rudykh17},
 \begin{equation}
{I}_1=\tr{\vec{c}}, \quad
{I}_2=\tr({\vec{c}^{-1}}),\quad
{I}_4= \vec E_L\vec\cdot \vec E_L,
\quad
{I}_5=\vec E_L\cdot\vec c^{-1}\vec E_L,
\quad
{I}_6=\vec E_L \cdot\vec c^{-2} \vec E_L,
\end{equation}
where $\vec{c}=\vec{F}^T\vec{F}$ is the right Cauchy-Green deformation tensor. 

In the appendix we present results for general materials, where in all generality $\Omega$ can be written as $\Omega=\Omega(I_1,I_2,I_4,I_5,I_6)$. In the main text we specialize the results to the \emph{Gent ideal dielectric} \citep{Gent96, Huang12a}, which exhibits the snap-through response. Its free energy is 
 \begin{equation}
\Omega_G=-\dfrac{\mu J_m}{2}\ln\left(1-\dfrac{I_1-3}{J_m}\right)-\dfrac{\varepsilon}{2}I_5,
\end{equation}
where $\mu$ is the initial shear modulus in the absence of electric field (in Pa), $J_m$ is the stiffening parameter (dimensionless) and $\varepsilon$ is the permitivity (in F/m).
When $J_m \to \infty$, the \emph{neo-Hookean ideal dielectric} \citep{ZhSu07} is recovered, 
\begin{equation} \label{nH}
\Omega_\text{nH} = \dfrac{\mu}{2}\left(I_1-3\right)-\dfrac{\varepsilon}{2}I_5,
\end{equation}
but that model does not provide snap-through loading behavior.

We call $H$ the initial thickness of the plate. 
We apply a voltage $V$ on the faces of the plate. 
Then the only non-zero component of the Lagrangian electric field is $E_{L2}=V/H$, which we call $E_0$. 
We call $x_1,x_3$ the in-plane Eulerian principal axes and $x_2$ the transverse axis, so that we have the following principal stretches and electric field components,
 \begin{equation}\label{equi-biaxial}
\lambda_1=\lambda_3=\lambda,
\quad
\lambda_2=\lambda^{-2},
\qquad
E_1=E_3=0,
\quad
E_2=\lambda^{2}E_0.
\end{equation}
Note that in the appendix we give results for a bi-axially stretched plate, when $\lambda_1$ is not necessarily equal to $\lambda_3$.

Introducing the function $\omega=\omega(\lambda,E_0)$ as
\begin{equation}\label{energy1}
\omega=\Omega(2\lambda^2+\lambda^{-4},2\lambda^{-2}+\lambda^{4},E_0^2,\lambda^4E_0^2,\lambda^8E_0^2),
\end{equation}
for a plate with no mechanical traction applied on the faces $x_2=\pm h/2$, where $h$ is the thickness of the deformed plate, we find the following compact expression for the equi-biaxial nominal stress component required to maintain the deformation \citep{Ogden14}:
\begin{equation} \label{s}
s=\dfrac{1}{2}\frac{\partial\omega}{\partial\lambda},
\end{equation}
where $s$ is the nominal stress applied along the in-plane directions. For example, for the Gent ideal dielectric we have \citep{lu12}
\begin{equation}
s_G = \mu\frac{\lambda-\lambda^{-5}}{1-(2\lambda^2+\lambda^{-4}-3)/J_m}-\varepsilon\lambda^3E_0^2,
\end{equation}
which we can easily invert to find the voltage $V_G =  E_0 H$.
Here, a non-dimensional measure of $V_G$ is
\begin{equation}
\overline{E}_0=\frac{V_G}{H\sqrt{\mu/\varepsilon}}=\frac{E_0}{\sqrt{\mu/\varepsilon}}=\sqrt{\frac{\lambda^{-2}-\lambda^{-8}}{1-(2\lambda^2+\lambda^{-4}-3)/J_m}-\lambda^{-3}\bar{s}},
\end{equation}
where $\bar s=s_G/\mu$ is a non-dimensional measure of stress.
We use this formula to plot the non-dimensional loading curves of Figure \ref{fig:snap1}, as well as the curve for the onset of snap-through, corresponding to $d\overline E_0/d\lambda=0$.

Note that the electric displacement vector $\vec D = (0,D,0)$, with $D$ being the only non-zero component, is related to the electric field through the formula
\begin{equation} \label{D}
D = -\lambda_1^{-1} \lambda_3^{-1}\dfrac{\partial \omega}{\partial E_0}.
\end{equation} 

\begin{figure}[ht!]
\centering
\includegraphics[width=0.9\textwidth]{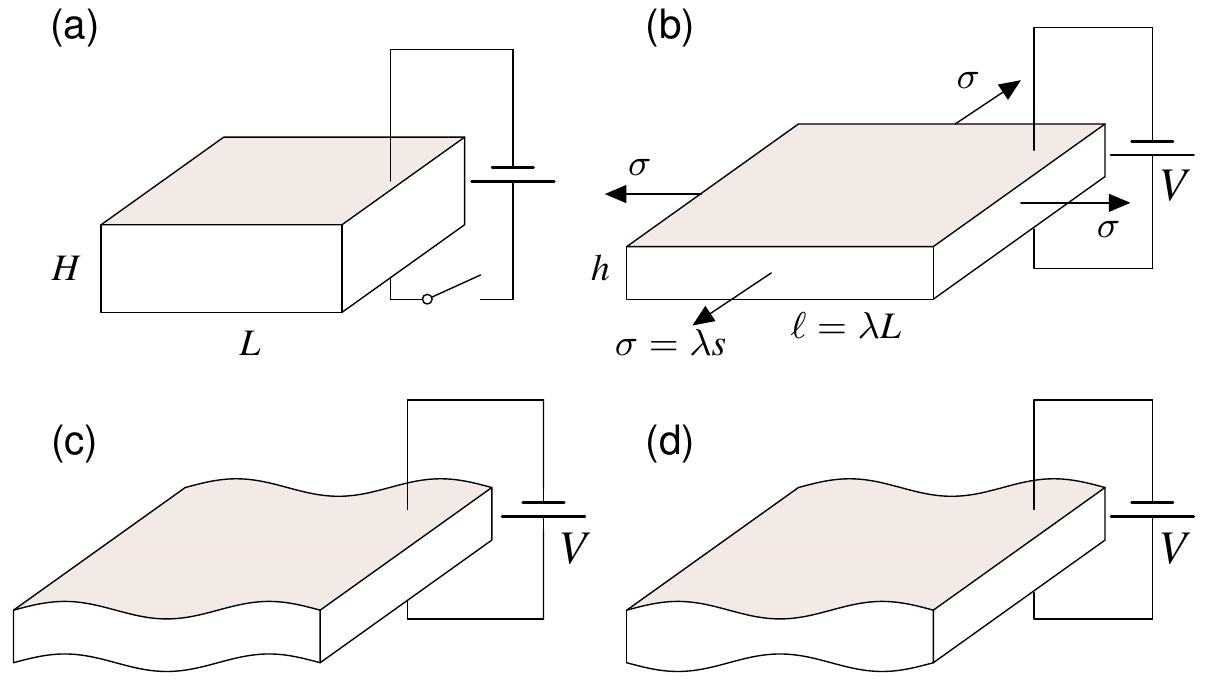}
\caption{
{\footnotesize
When put under voltage $V$ and/or stress $\sigma$, a rectangular plate made of a soft dielectric and with faces covered by compliant electrodes deforms homogeneously (a)-(b). It can even deform so severely as to loose its stability and buckle into antisymmetric (c) or symmetric (d) modes of wrinkles.
}}
\label{fig:sketch}
\end{figure}

In this paper we study the possibility of homogeneously deformed plates buckling inhomogeneously as sketched in Figure \ref{fig:sketch}.
Typically, we find that the onset of these buckling modes is governed by a \emph{dispersion equation} relating the critical stretch $\lambda_\text{cr}$ to the ratio of the plate thickness $H$ by the wavelength of the wrinkles $\mathcal L$.
In effect, we find that it can be factorised into the product of a dispersion equation for antisymmetric wrinkles and one for symmetric wrinkles, see Figures \ref{fig:sketch}(c)-(d). 
Note that the results are also valid for the case of a plate of finite width (initial width: $L$, current width: $\ell = \lambda L$) confined between two lubricated rigid walls, as in the paper by \cite{Yang17}: in that case $\mathcal L = \ell / m$, where $m$ is the  number of  wrinkles.

\section{Small-amplitude wrinkles}
\label{section3}


We now linearize the governing equations and boundary conditions in the neighborhood of the large electro-elastic deformation.

We introduce the following fields: $\vec u$, the small-amplitude mechanical displacement; $\dot T_{2i}$, the incremental mechanical traction on the planes $x_2= $const.; and $\vec{\dot D}_L$, ${\vec{\dot E}}_L$, the incremental electric displacement and electric field, respectively. 
These fields are functions of $\vec x$, the position vector in the current (actuated) configuration \citep{Ogden2014}.

We focus on \emph{two-dimensional wrinkles}, and thus take the fields to be functions of $x_1,x_2$ only.
This leads to (not shown here) $u_3=\dot{T}_{3i}=\dot{E}_{L3}=\dot{D}_{L3}=0$.
Because the updated, incremental version of the Maxwell equation $\Curl {\vec E_L}=\vec 0$ is $\curl \vec{\dot E}_L = \vec 0$ \citep{Ogden10}, we can introduce the electric potential $\varphi$ by
\begin{equation}
\dot{{E}}_{L1}=-\partial{\varphi}/\partial{x_1},
\qquad
\dot{{E}}_{L2}=-\partial{\varphi}/\partial{x_2}.
\end{equation}
We then seek solutions with sinusoidal shape along $x_1$ and amplitude variations along $x_2$, in the form
\begin{equation}
\left\{ {{u}_{1}},{{u}_{2}},{{{\dot{D}}}_{L2} ,{{{\dot{T}}}_{21}},{{{\dot{T}}}_{22}}},\varphi \right\}=  
\Re \left\{\left[k^{-1} U_1, k^{-1} U_2,\ii \Delta,\ii \Sigma_{21},\ii \Sigma_{22}, k^{-1} \Phi\right] e^{\ii kx_1}\right\},
\end{equation}
where  $U_1$, $U_2$, ${{\Delta }}$, ${{\Sigma }_{21}}$, ${{\Sigma }_{22}}$, $\Phi$ are functions of $k{{x}_{2}}$ only and $k =2\pi/\mathcal L$ is the wavenumber.

Our main result is that the governing equations can be put in the form
\begin{equation}\label{stroh}
\vec{\eta}'=\ii\vec{N}\vec{\eta} ,
\end{equation}
where 
\begin{equation}
\label{eta0}
\vec{\eta} ={{\left[ \begin{matrix}
   U_1 & U_2 & {{\Delta }}  & {{\Sigma }_{21}} & {{\Sigma }_{22}} & \Phi  \\
\end{matrix} \right]}^{T}}=\left[\begin{matrix}
\vec{U} & \vec{S}
\end{matrix} \right]^{T},
\end{equation}
 is the Stroh vector, the prime denotes differentiation with respect to $k{{x}_{2}}$.
 Here $\vec U={{\left[ \begin{matrix}
   U_1 & U_2 & {{\Delta }}  \\
\end{matrix} \right]}^{T}}$, $\vec S={{\left[ \begin{matrix}
 {{\Sigma }_{21}} & {{\Sigma }_{22}} & \Phi  \\
\end{matrix} \right]}^{T}}$ are the generalised displacement and traction vectors, respectively, and $\vec N$ is the Stroh matrix.
In the appendix we show that $\vec N$ has the following block structure 
\begin{equation}
\vec{N}=\left[ \begin{matrix}
   {{\vec{N}}_{1}} & {{\vec{N}}_{2}}  \\
   {{\vec{N}}_{3}} & \vec{N}_{1}  
\end{matrix} \right],
\end{equation}
where the  ${{\vec{N}}_{i}}$ $(i=1,2,3)$ are real symmetric. 
We find that these $3\times 3$ sub-matrices are 
\begin{align} \label{stroh-matrix}
 \vec{N}_1 = \left[ \begin{matrix}
   0 & -1 & 0  \\[4pt]
   -1 & 0 & 0  \\[4pt]
   0 & 0 & 0  
\end{matrix} \right],
\quad
\vec{N}_{2} = \left[ \begin{matrix}
  \dfrac{ 1}{c} & 0 & \dfrac{d}{c}  \\[10pt]
   0 & 0 & 0  \\[10pt]
   \dfrac{d}{c} & 0 & \dfrac{d^2}{c} - f 
\end{matrix} \right], 
\quad
  \vec{N}_3 = \left[ \begin{matrix}
   \dfrac{e^2}{g} - 2(b+c) & 0 & - \dfrac{e}{g}  \\[10pt]
   0 & c- a & 0  \\[4pt]
   -\dfrac{e}{g} & 0 & \dfrac{1}{g}  
\end{matrix} \right],
\end{align} 
where $a$, $b$, $c$, $d$, $e$, $f$, $g$ are electro-elastic moduli.
Their general expression in terms of the free energy density $\Omega$ is given in the appendix.
For equi-biaxial deformations, they read
\begin{align}
a &= 2\left[\lambda^2 (\Omega _1 + \lambda ^2 \Omega_2) + \lambda^4 \left( \Omega_5+(2\lambda^4 + \lambda^{-2})\Omega_6 \right)  E_0 ^2\right], \notag \\[12pt]
b & = 2 \left\lbrace (\lambda^{-4}-\lambda^2) \left[ (\lambda^{-4}-\lambda^2)(\Omega_{11} +2\lambda^2 \Omega_{12} + \lambda^4 \Omega_{22})  \right. \right. \notag \\
 & \qquad \qquad\qquad \left.   -2\lambda^4 (\Omega_{15} + 2\lambda^4 \Omega_{16}+ \lambda^2 \Omega_{25} + 2\lambda^6 \Omega_{26}) E_0 ^2\right]
 \notag \\
 & \qquad \qquad\qquad  \left. + \lambda^8 (\Omega_{55} +4\lambda^4 \Omega_{56} + 4\lambda^8 \Omega_{66})E_0 ^4 \right\rbrace 
 \notag \\
& \qquad \qquad\qquad  +   (\lambda^{2} + \lambda^{-4})(\Omega_1 + \lambda^2 \Omega_2) + \lambda^4 \left[ \Omega_5 + 2(3\lambda^4 - \lambda^{-2})\Omega_6 \right]E_0 ^2, \notag \\[12pt]
c & = 2\lambda^{-2}\left[ \lambda^{-2}(\Omega_1 + \lambda^2 \Omega_2 ) + \lambda^4 \Omega_6 E_0^2 \right], 
\notag \\[12pt]
d& = -2 \lambda^2 \left[ \Omega_5 + (\lambda^4 + \lambda^{-2}) \Omega_6 \right] E_0, 
\notag \\[12pt]
e &= 4 \lambda^2\left[ (\lambda^{-4} - \lambda^2)(\lambda^{-4} \Omega_{14} + \Omega_{15} + \lambda^4 \Omega_{16} + \lambda^{-2} \Omega_{24} + \lambda^2 \Omega_{25} + \lambda ^6 \Omega_{26}) \right.
 \notag \\
& \qquad \left. - \lambda^4 (\lambda^{-4}\Omega_{45} +2\Omega_{46} + \Omega_{55} +3\lambda^4 \Omega_{56} +2\lambda^8 \Omega_{66})E^2 _0 -(\Omega_5 + 2\lambda^4 \Omega_6) \right] E_0,
 \notag \\[12pt]
f & = 2(\lambda^2 \Omega_4 + \Omega_5 + \lambda ^{-2} \Omega_6),
 \notag \\[12pt]
g & = 4 \lambda^4\left[ \lambda^{-8} \Omega_{44} + 2\lambda^{-4} \Omega_{45} + 2\Omega_{46} +\Omega_{55} + 2\lambda^4 \Omega_{56} + \lambda^8 \Omega_{66} \right] E_0 ^2 \notag \\
& \qquad \quad+2(\lambda^{-4} \Omega_4 + \Omega_5 + \lambda^4 \Omega_6), \label{eqb_const}
\end{align}
where $\Omega_i = \partial \Omega / \partial I_i$ and $\Omega_{ij} = \partial ^2 \Omega / \partial I_i \partial I_j$.

Specializing to the Gent ideal dielectric, we find
\begin{align} 
& a = \mu (2\lambda^2 \overline W' - \lambda^4 \overline E_0^2), 
&&
 c = 2 \mu \lambda^{-4} \overline W',
 && 
  2b = 4 \mu (\lambda^{-4}-\lambda^2)^2 \overline W''+a+c,\\
& d= \sqrt{\mu \varepsilon}\lambda^2 \overline E_0,
&& e = 2 d,
&& f = g = - \varepsilon,
\end{align} 
where 
\begin{equation}
\overline W'= \frac{1}{2\left[1 - (2\lambda^2+\lambda^{-4}-3)/J_m\right]},
\qquad
\overline W''=\frac{1}{2J_m\left[1 - (2\lambda^2+\lambda^{-4}-3)/J_m\right]^2}.
\end{equation}

The Stroh equation must be solved subject to the incremental boundary conditions on the faces of the plate of no mechanical traction and no electrical field, i.e.,
\begin{equation}
\vec S(-kh/2)=\vec 0,
\qquad
\vec S(kh/2)=\vec 0.
\end{equation}

Now because the Stroh matrix is constant, the resolution of \eqref{stroh} is straightforward. 
It reduces to an eigenvalue problem, yielding a complete set of six linearly independent eigensolutions with exponential variations in $x_2$.
Then the boundary conditions give a linear $6 \times 6$ homogeneous system of equations for the six unknowns $U_1(\pm h/2), U_2(\pm h/2), \Delta(\pm h/2)$, for which the determinant must be zero: this is the dispersion equation.

Using the usual matrix manipulations of plate acoustics (see  \cite{Nayfeh95} for instance), the six exponential solutions can be decoupled into two sets of three (hyperbolic) trigonometric solutions, one corresponding to antisymmetric modes, the other to symmetric modes, see sketches in Figure \ref{fig:sketch}.
The dispersion equation itself factorises into two corresponding equations.
In the appendix we give those equations for an arbitrary tri-axial pre-stretch, for materials with free energies of the forms
\begin{equation}
\Omega = W(I_1) - \dfrac{\varepsilon}{2} I_5, \qquad
\Omega = \dfrac{\mu(1-\beta)}{2}(I_1-3) +\dfrac{\mu \beta}{2}(I_2-3) - F(I_5), \label{MR-F} 
\end{equation}
where $W$ and $F$ are arbitrary functions (note that the Gent ideal dielectric belongs to the first type), and $0 \le \beta \le 1$ is a constant.

For the Gent ideal dielectric under an equi-biaxial pre-stretch \eqref{equi-biaxial}, we find the following \emph{explicit dispersion equation for the anti-symmetric wrinkles}:
\begin{multline} \label{anti}
 2  \overline W' \left[p_1 (1 + p_2^2)^2 \tanh\left(\pi p_1 \lambda^{-2} H/\mathcal L \right) - p_2 (1 + p_1^2)^2\tanh\left(\pi p_2 \lambda^{-2}H/\mathcal L \right)\right]   \\[4pt]
 = ( p_2^2 - p_1^2) \lambda^{8} \overline E_0^2 \tanh \left( \pi \lambda^{-2}H / \mathcal L \right),
\end{multline}
where 
\begin{equation}
p_{1,2} = \dfrac{\lambda^{3}+1}{2}\sqrt{1+2(\lambda - \lambda^{-2})^2\dfrac{\overline W''}{\overline W'}} \pm  \dfrac{\lambda^{3}-1}{2}\sqrt{1 + 2(\lambda + \lambda^{-2})^2\dfrac{\overline W''}{\overline W'}}.
\end{equation}
For symmetric wrinkles, the dispersion equation is the same except that $\tanh$ is replaced with $\coth$ everywhere.

In the case of the neo-Hookean ideal dielectric \eqref{nH}, we take $J_m \to \infty$ and have $\overline W'=1/2$, $\overline W''=0$, $p_1=\lambda^3$, $p_2=1$, and the dispersion equations simplify to
\begin{equation} \label{Neo}
\left[\dfrac{\tanh(\lambda \pi H/\mathcal L)}{\tanh(\lambda^{-2}\pi H/\mathcal L)}\right]^{\pm 1} = 
\dfrac{(1 + \lambda^6)^2}{4\lambda^3} + \dfrac{\lambda^5(1 - \lambda^6)}{4}\overline E_0^2,
\end{equation}
where the $+1$ ($-1$) exponent corresponds to anti-symmetric (symmetric) wrinkles.

We now plot the  dispersion curves for the Gent ideal dielectric as the plate is loaded homogeneously by an increasing voltage.

\begin{figure}[ht!]
\centering
\includegraphics[width=\textwidth]{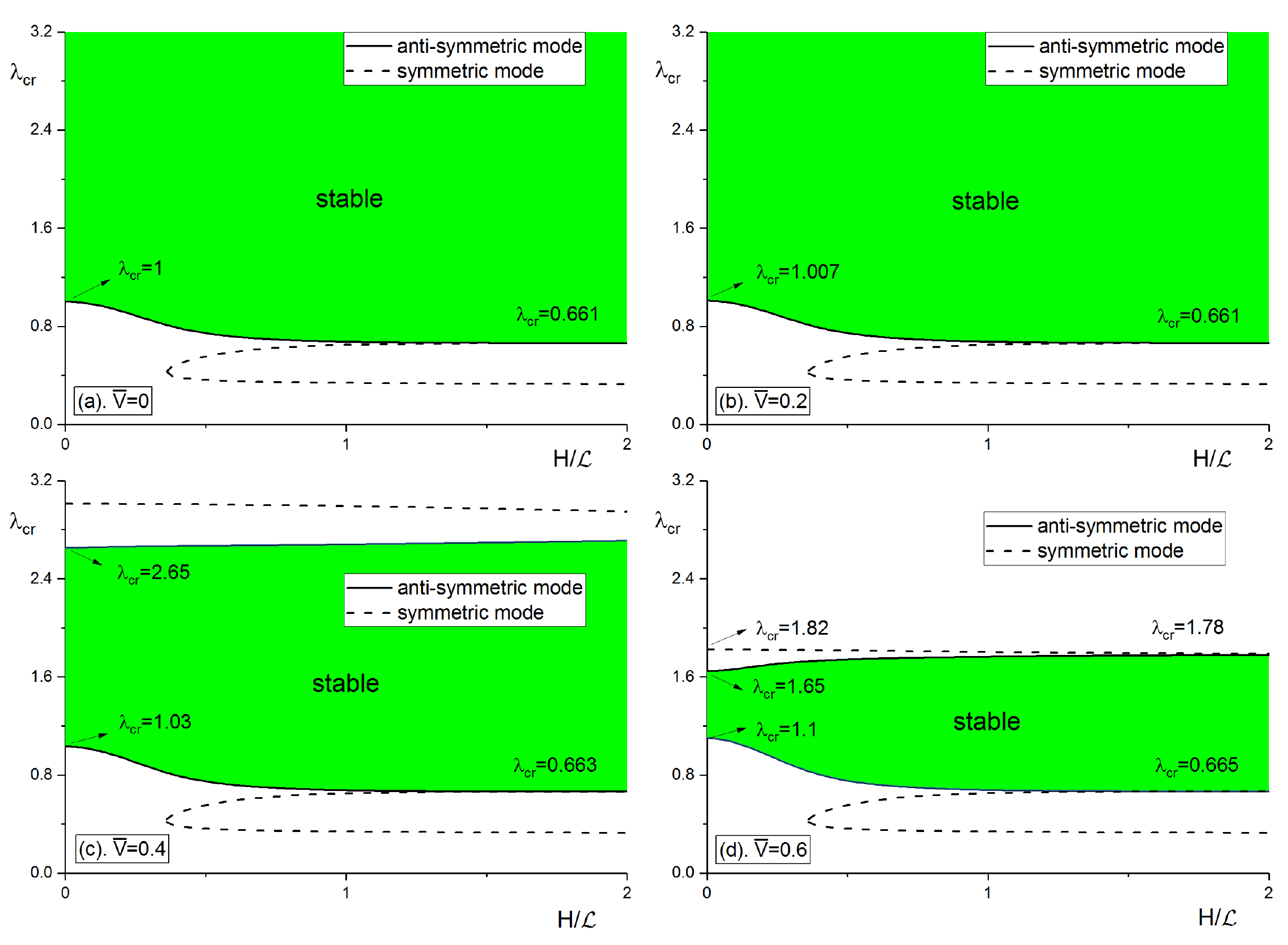}
\caption{
{\footnotesize
Dispersion curves for electrically loaded ($\overline E_0 = 0.0, 0.2, 0.4, 0.6$) dielectric plates: critical stretch ratio $\lambda_\text{cr}$ of compression (lower curves) and of extension (upper curves) against the initial thickness to wavelength ratio $H/\mathcal L$.
(a)-(b): For low voltages, the plate can wrinkle only in compression. (c)-(d): For higher voltages, the dielectric plate can wrinkle in extension, see the thick upper line, corresponding to anti-symmetric wrinkles. 
Then, the critical stretch is located between its limit values in the thin-plate ($H/\mathcal L \to 0$) and short-wave ($H/\mathcal L \to \infty$) limits.
}
}
\label{fig:disp}
\end{figure}

When $\overline E_0=0$, see Figure \ref{fig:disp}(a), we recover the purely mechanical case.
The lower/dashed (upper/full) curve corresponds to symmetric (anti-symmetric) buckling. 
We see that in extension ($\lambda>1$), the plate is always stable, whereas in contraction ($\lambda<1$), it buckles antisymmetrically, with $\lambda_\text{cr}\simeq 1$ when $H/\mathcal L$ is small (thin plate, long wavelength) and $\lambda_\text{cr}\simeq 0.661$ when $H/\mathcal L$ is large (thick plate, short wavelength). 
Note that here ``large" and ``thick" simply mean that the plate initial thickness $H$ is of the order of the wavelength $\mathcal L$.

When $\overline E_0=0.2$, see Figure \ref{fig:disp}(b), the landscape is the same, with the curves slightly shifted upwards.
The plate only buckles in contraction, a possibility that we rule out, because we are only interested in extending the plate through voltage (and possibly, pre-stretch).

When $\overline E_0=0.4$, see Figure \ref{fig:disp}(c), we see that the possibility of \emph{buckling in extension} has now emerged. 
The plate now buckles anti-symmetrically when $\lambda$ reaches a critical value $\lambda_\text{cr}$ between 2.65 (thin-plate limit) and 2.81 (short-wavelength limit), depending on the ratio $H/\mathcal L$.

Similarly, when $\overline{E}_0=0.6$,  Figure \ref{fig:disp}(d) shows that the plate buckles anti-sym\-me\-tri\-cal\-ly in extension when $\lambda$ reaches a critical value $\lambda_\text{cr}$ between 1.65 (thin-plate limit) and 1.78 (short-wave limit), depending on the ratio $H/\mathcal L$.

From this rapid analysis, we conclude that it is unnecessary to study the dispersion equation in detail for the Gent ideal dielectric, and that the thin-plate and short-wave limits suffice to find global, wavelength-independent critical stretches of wrinkling in extension.

 \section{Thin-plate and short-wave instabilities}
 \label{section4}
 

In the previous section we saw that put under a sufficiently large voltage, a dielectric plate buckles anti-symmetrically in extension.
Depending on the ratio $H/\mathcal L$ of thickness to wavelength, the plate buckles at a critical stretch located in between a lower bound, corresponding to the limit for thin plates $H/\mathcal L \to 0$ and an upper bound, the limit for short wavelengths $H/\mathcal L \to \infty$.
In this section we present explicit expressions for these two limits.

First, the \emph{thin-plate limit} can be found with the asymptotic behavior of $\tanh$ as its argument is small in the dispersion equation \eqref{anti}. 
However, we can in fact give the thin plate limit in the most general case.
Using the results of \cite{Shuv00}, it is easy to show that the buckling condition when $H/\mathcal L \to 0$ is simply
\begin{equation}
\det \vec N_3 = 0,
\end{equation}
where $\vec N_3$ is the Stroh sub-matrix given in Equation \eqref{stroh-matrix}.
It factorises to give $(a-c)(b+c)=0$, and anti-symmetric buckling corresponds to $a-c=0$. 
Combining Equations \eqref{energy1} and \eqref{s}, we find that $a-c=(\lambda/2) \partial\omega/\partial\lambda$, so that anti-symmetric buckling is equivalent to
\begin{equation} \label{thin}
\frac{\partial\omega}{\partial\lambda} =0.
\end{equation}
Comparing with Equation \eqref{s}, we see that in general, the loading $\overline E_0-\lambda$ curve with no pre-stress ($\overline s=0$) is in fact the buckling limit for plates of vanishing thickness. 
Here the equation reads
\begin{equation} \label{thin-G}
 \frac{\lambda^{-2} - \lambda^{-8}}{1-(2\lambda^2+\lambda^{-4}-3)/J_m} = \overline E_0^2.
\end{equation}

Next the \emph{short-wave limit} is found from Equation \eqref{anti} by replacing $\tanh$ with 1, its value as $H/\mathcal L \to \infty$. 
After some re-arrangement, we find that it reads as
\begin{equation}\label{short}
2\lambda(\lambda^9 + \lambda^6 + 3 \lambda^3 - 1) \overline W' + 4(\lambda^6-1)^2 \overline W'' = \lambda^9 (1+\lambda^3)\overline E_0^2\sqrt{1+2(\lambda-\lambda^{-2})^2\dfrac{\overline W''}{\overline W'}}.
\end{equation}
For example, when $\overline E_0=0.6$ (and $J_m=97.2$) we find that the root to this equation is $\lambda_\text{cr} = 0.665$, as reported on Figure \ref{fig:disp}(d).
Note that the purely elastic case ($\overline E_0 =0$) is consistent with the surface stability criterion of an elastic Gent material \citep{DeSc04}, giving $\lambda_\text{cr} = 0.661$ here.
For the neo-Hookean ideal dielectric of Equation \eqref{nH}, the equation simplifies to 
\begin{equation}
\lambda^9 + \lambda^6 + 3 \lambda^3 - 1 = \lambda^8 (1+\lambda^3)\overline E_0^2,
\end{equation}
Note that \cite{Ogden14, Ogden2014} studied the surface instability of an ideal neo-Hookean dielectric, but it was charge-controlled instead of voltage-controlled as here.

Now the two equations \eqref{thin-G} and \eqref{short} delineate a region in the $\overline E_0 - \lambda$ landscape of Figure \ref{fig:snap1} where a given plate is going to buckle in extension when subject to a sufficiently large $\overline E_0$. 
The precise value of the corresponding $\lambda_\text{cr}$ depends on $H/\mathcal L$ but the region between the two curves is narrow enough to draw general conclusions.

\begin{figure}[h!]
\centering
\includegraphics[width=0.9\textwidth]{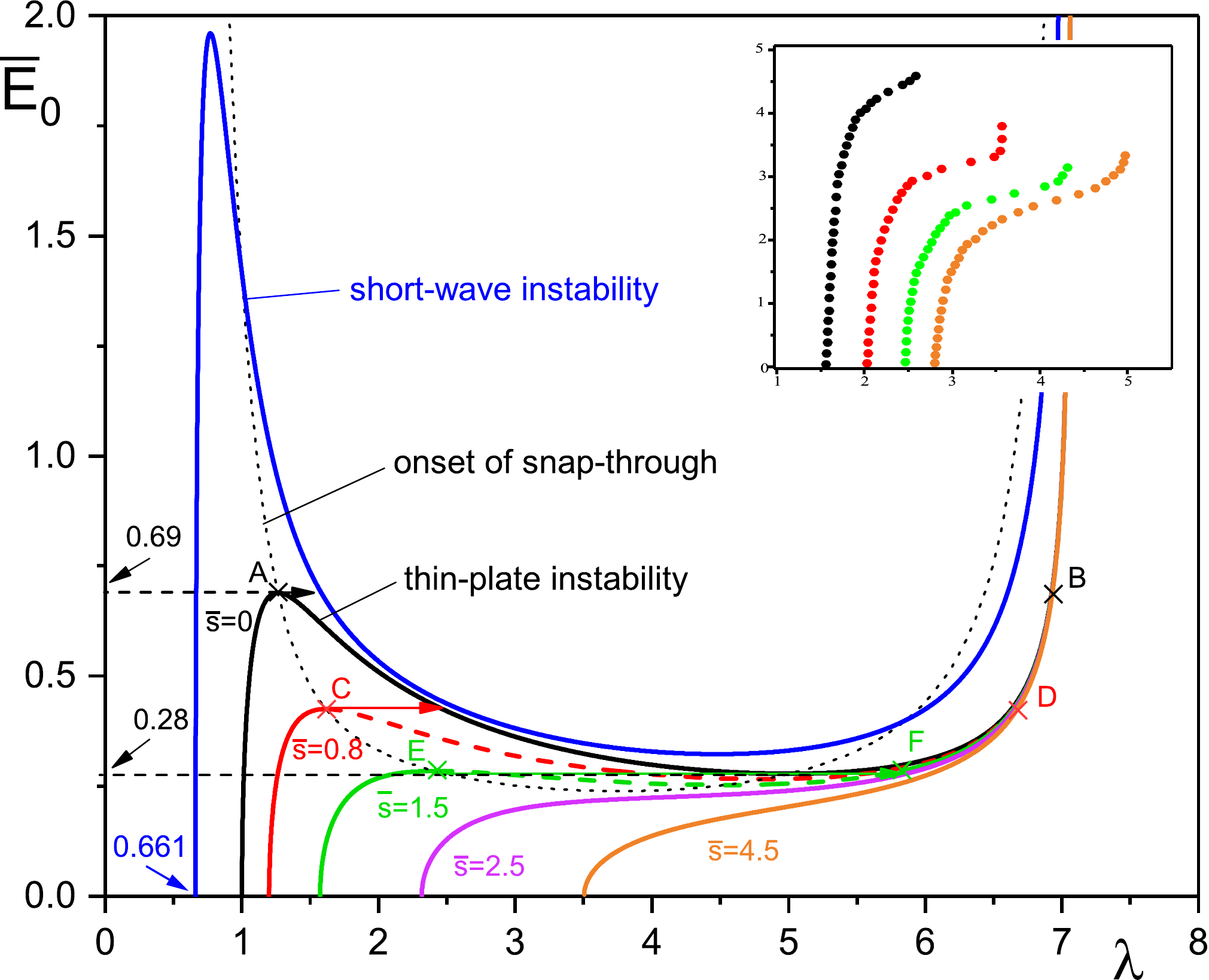}
\caption{
{\footnotesize
How the snap-through actuation is counter-acted by plate instabilities. Solid lines are the voltage-stretch curves for homogeneous loading at different levels of pre-stress ($\bar s=0,0.8,...,4.5$), when the plate is modeled by the Gent ideal dielectric (here $J_m=97.2$, \citep{Gent96, DorfOgden14}). 
Their intersection with the black dashed line shows where a snap-through should start.
However, the non-prestressed plate ($\overline s = 0$) will meet the buckling zone found between the thin-plate and the short-wave limit curves on its way from A to B. 
Similarly for the $\bar s=0.8$ pre-stressed plate. 
Only the $\bar s = 1.5$ pre-stressed plate might be able to experience a snap-through transition from E to F.
Inset: Experimental voltage (kV)$-$stretch data from \cite{Huang12}.}}
\label{fig:snap2}
\end{figure}

In Figure \ref{fig:snap2} we plot the loading curves for the same Gent ideal dielectric ($J_m=97.2$) used to generate the plots of Figure \ref{fig:snap1}, together with the curves for the thin-plate and short-wave limits.
We see that the snap-through scenario is not going to unfold for the $\overline s=0$ curve: as soon as $\overline E_0$ reaches its maximum (A) and the plate starts expanding with voltage remaining fixed at that value, we enter the buckling zone between the thin-plate and the short-wave limits. 
The same is true for the $\overline s = 0.8$ pre-stressed plate, as again, the snap-through from C to D hits the buckling zone and cannot be completed. 
Only the  $\overline s = 1.5$ pre-stressed plate might be able to achieve a snap-through from E to F without buckling.

For plates subject to sufficiently large pre-stress ($\overline s=2.5, 4.5$),  the snap-through, thin-plate and  short-wave instabilities are avoided, and the plates can deform until they fail by electrical breakdown.

In Figure \ref{fig:disp}, we saw that plates  buckle in compression only when $\overline E_0$ is small (e.g., $\overline E_0=0.0, 0.2$, see Figures \ref{fig:disp}(a),(b)), and then in compression \emph{and} in tension for sufficiently large voltage (e.g., $\overline E_0=0.4, 0.6$, see Figures \ref{fig:disp}(c),(d)). 
The critical voltage where the buckling in tension is first expressed is determined by 
\begin{equation} \label{critical1}
\frac{\partial\overline E_0}{\partial\lambda} =0, \quad
\text{where} \quad
\overline E_0=\sqrt{\dfrac{\lambda^{-2}-\lambda^{-8}}{1-(2\lambda^2+\lambda^{-4}-3)/J_m}}
\end{equation}
corresponds to the  $\overline E_0-\lambda$ loading curve with no pre-stress ($\overline s=0$).

In our calculation ($J_m=97.2$), Equation \eqref{critical1} has two real roots $\overline E_{0}=0.69$ and $\overline E_{0}=0.28$. 
The former corresponds to the voltage at point A (onset of snap-through at the local maximum of the curve) and the latter  to the voltage at the local minimum of the curve, see Figure \ref{fig:snap2}. 

Finally, we inserted experimental results in   Figure \ref{fig:snap2} on actual voltage-stretch curves due to  \cite{Huang12}.
They do show that plates can be stretched by voltage a little bit further than the onset of snap-through suggests, by going beyond the maximum of the curve, and that pre-stretch allows for further absolute stretch of the plate by voltage, although the relative gain is affected by the pre-stretch, see  \cite{Huang12} for a more detailed discussion.



\section{Further results}



\subsection{Plate wrinkling for dielectrics with polarization saturation}

\begin{figure}[hb!]
\centering
\includegraphics[width=.9\textwidth]{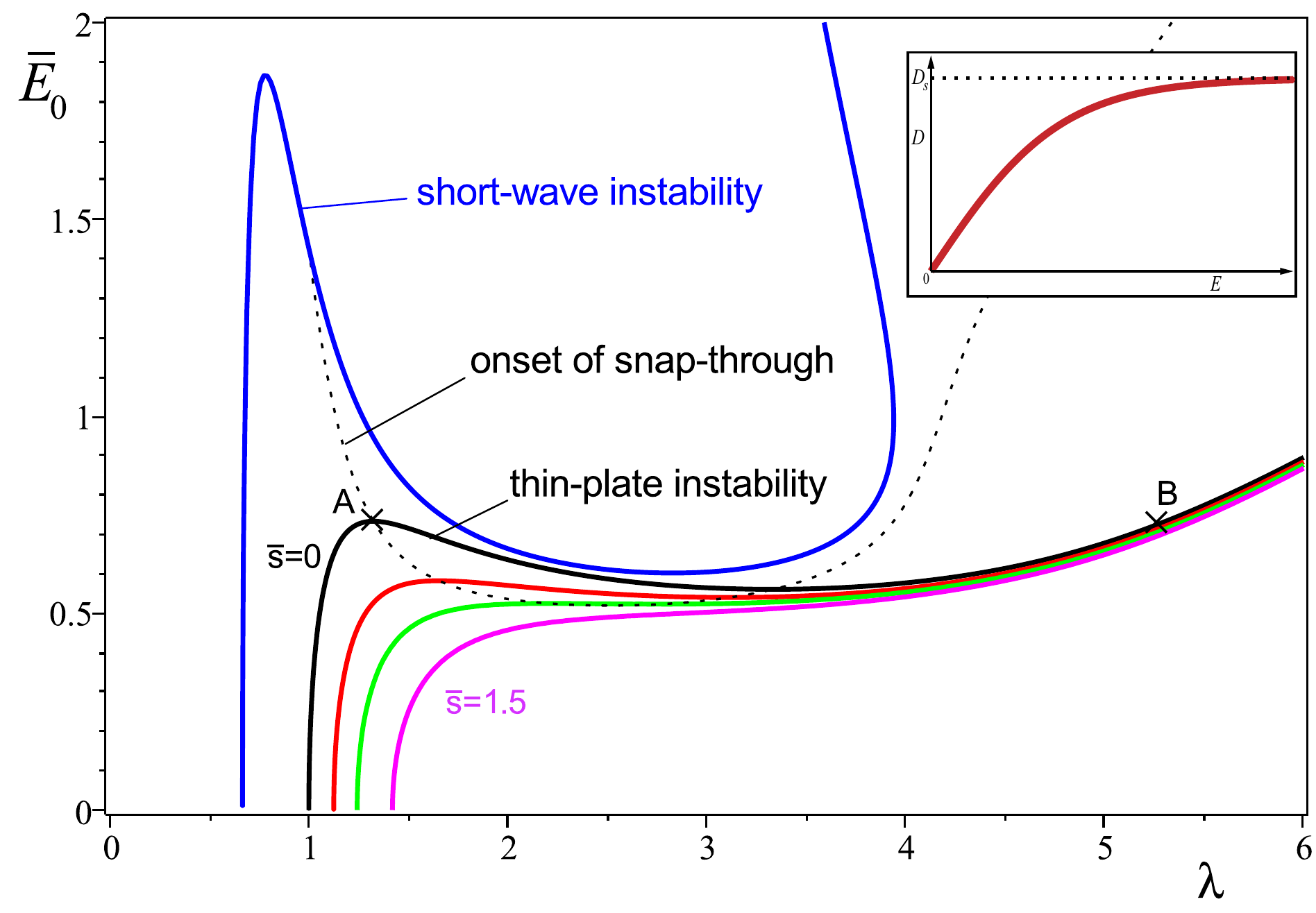}
\caption{
{\footnotesize
A dielectric with polarisation saturation: Solid lines are the voltage-stretch curves for homogeneous loading at different levels of pre-stress ($\bar s=0,0.6,1.0,1.5$). 
The non-prestressed plate ($\overline s = 0$) will meet the buckling zone found between the thin-plate and the short-wave limit curves on its way from A to B. 
The pre-stressed plates might be able to avoid buckling, depending on their thickness.
Inset: Polarisation saturation of the electric displacement with the electric field for the corresponding model \citep{Liu12}.}}
\label{fig:polarisation}
\end{figure}

Many dielectrics exhibit the phenomenon of \emph{polarization saturation}, in the sense that the electric displacement $D$ increases monotonically with the electric field $E$ but with an asymptotic upper bound $D_s$, say \citep{LiSu11, Li12, Liu12}.
This characteristic can be captured by the following form of energy density,
\begin{equation}
\Omega = \dfrac{\mu(1-\beta)}{2}(I_1-3) +\dfrac{\mu \beta}{2}(I_2-3) - \dfrac{D_s^2}{\varepsilon}\ln\left(\cosh\left(\varepsilon \dfrac{\sqrt{I_5}}{D_s}\right)\right), \label{saturation}
\end{equation}
where $D_s>0$ and $0 \le \beta \le 1$ are constants.
Then the electric displacement $D$ is related to the electric field $E$ through (see Equation \eqref{D} and Figure \ref{fig:polarisation}),
\begin{equation}
D = D_s \tanh(\varepsilon E/D_s).
\end{equation}

The nominal stress required to effect an equi-biaxial stretch with a transverse electrical field is found from Equation \eqref{s} as
\begin{equation}
s = \mu(1-\beta)(\lambda - \lambda^{-5}) + \mu \beta (\lambda^3-\lambda^{-3}) - \lambda D_s E_0 \tanh\left(\lambda^2 {\varepsilon E_0}/{D_s}\right).
\end{equation}
We introduce the following quantities
\begin{equation}
\overline s = s/\mu, \qquad 
\overline E_0 = E_0\sqrt{\varepsilon/\mu}, \qquad
\overline D_s = D_s/\sqrt{\mu \varepsilon},
\end{equation}
to obtain the non-dimensional version of this equation as
\begin{equation} \label{s-pol}
\overline s = (1-\beta)(\lambda - \lambda^{-5}) + \beta (\lambda^3-\lambda^{-3}) - \lambda \overline D_s \overline  E_0 \tanh\left(\lambda^2 {\overline  E_0}/{\overline D_s}\right).
\end{equation}
For a given level of pre-stress $\overline s$, it gives an implicit relationship between the voltage and the stretch, which we solve to plot the $\overline E_0-\lambda$ curves of Figure \ref{fig:polarisation}. 
For these plots we took $\beta=0.2$ and $\overline D_s= 4\sqrt{5}$ (corresponding to $k=1/4$ and $D_s/\sqrt{C_1\varepsilon}=10$ in the paper by \cite{Liu12}).
We also plot the curve for the onset of snap-through, found by differentiating implicitly  Equation \eqref{s-pol} and taking $d\overline E_0/d\lambda=0$.

We also note that, as outlined above, the \emph{thin-plate limit} corresponds to this equation in the absence of pre-stress, i.e., when $\overline{s}=0$.
Since the energy density \eqref{saturation} is of the form \eqref{MR-F}$_2$, we can make use of the results found using this choice of energy (see appendix, \eqref{short-wave-f}), to obtain the \emph{short-wave limit} as
 \begin{equation}
p_3\left[\lambda^{-2} (1-\beta) + \beta \right] (\lambda^3 + 1 + 3\lambda^{-3} - \lambda^{-6}) - (\lambda + \lambda^{-2}){\overline E_0}{\overline D_s} \tanh \left(\lambda^2 {\overline E_0}/{\overline D_s}\right)= 0, \label{short-wave-sat}
\end{equation}
where \begin{equation}
p_3 = \sqrt{\frac{{\overline D_s}}{2\lambda^2 {\overline E_0}} \sinh \left( \frac{2 \lambda^2 {\overline E_0}}{\overline D_s} \right)},
\end{equation}
is the imaginary part of the third eigenvalue of the corresponding Stroh matrix.

We plotted the thin-plate limit and short-wave limit curves in Figure \ref{fig:polarisation}. 
We can see that when the material is not under mechanical pre-stress ($\overline {s} =0$), snap-through does not occur, as the buckling zone between the thin-plate and short-wave instabilities is reached before it can be completed. However, we can also see that the plate may avoid buckling when it is pre-stressed, and potentially achieve snap-through, depending on the thickness of the material.


\subsection{Correction to the thin-plate buckling equation}


Finally, we can exploit the exact dispersion equations to establish approximations to the dispersion equations when the plate is thin.
For this exercise, we specialise the analysis to the \emph{Mooney-Rivlin ideal dielectric} model, with free energy density
\begin{equation}
\Omega = \dfrac{\mu(1-\beta)}{2}(I_1-3) +\dfrac{\mu \beta}{2}(I_2-3) - \dfrac{1}{2\varepsilon} I_5.
\end{equation}

In this case, the dispersion equation for anti-symmetric wrinkles (the first to appear) reads
\begin{equation} \label{mooney1}
\dfrac{\tanh(\lambda \pi H/\mathcal L)}{\tanh(\lambda^{-2}\pi H/\mathcal L)}= 
\dfrac{(\lambda^2\beta+1-\beta)(1+\lambda^6)^2+\overline E_0^2 \lambda^8(1-\lambda^6)}{4\lambda^3\left[1+\beta(\lambda^2-1)\right]},
\end{equation}
which reduces to Equation \eqref{Neo} for $\beta=0$.
At the zero-th order in $H/\mathcal L$, we have the thin-plate equation \eqref{thin}, here:
\begin{equation}
\overline{E_0}^2 = (1- \beta + \beta \lambda^2)(\lambda^{-2}-\lambda^{-8}).
\end{equation}
This equation has one root (low values of $\overline{E} _0$)  or two roots (higher $\overline{E} _0$)  for $\lambda$, which we call $\lambda_0$.

The next order in  $H/\mathcal L$ is order two. 
With some manipulations of Equation \eqref{mooney1}, we find the following correction,
\begin{equation} \label{MR-euler}
\lambda = \lambda_0 - \dfrac{2}{3}\left[ \dfrac{\lambda_0^{3} (1-\beta + \beta \lambda_0^2)}{3\beta \lambda_0^2 + (1-\beta)(4-\lambda_0^6)}\right]\left(\pi {H}/{\mathcal L}\right)^2.
\end{equation}
This expression is valid  for both the smallest root of the thin-plate equation (corresponding to buckling in compression) and the largest root (buckling in extension under a large voltage), when it exists.
That latter case is a departure from the purely elastic case ($\lambda_0 \equiv 1$), where there is no loss of stability in extension \citep{Beatty1998}.
For the neo-Hookean ideal dielectric \eqref{nH}, we take $\beta=0$ and the expression reduces to
\begin{equation} \label{lambda-nH}
\lambda = \lambda_0 - \dfrac{2}{3}\left[ \dfrac{\lambda_0^{3} }{4-\lambda_0^6}\right]\left(\pi {H}/{\mathcal L}\right)^2.
\end{equation}

Finally, $\lambda_0=1$ in the purely elastic case, and from Equation \eqref{MR-euler} we recover the  Euler solution for the buckling of a slender column under equi-biaxial load: $\lambda = 1 - (2/9) (\pi H/\mathcal L)^2$, see \cite{Beatty1998} for the connection with the classical formula of the corresponding critical end thrust (see also \cite{Yang17}). 

This type of expansion in $H/\mathcal L$ can be performed for any free energy by using the Stroh matrix, see \cite{Shuv00} for details. 
It allows us to link our stability analysis to that based on the Hessian criterion.
That stability criterion is based on minimising the free energy once it has been expanded in terms of the plate thickness up to the first power \citep{Zurlo17}.
It corresponds to the onset of snap-through criterion, $d\overline E_0/d\lambda=0$.
For instance, take the neo-Hookean ideal dielectric \eqref{nH}: in the absence of a pre-stress, the onset of snap-through/Hessian criterion occurs when \citep{ZhSu07}
\begin{equation}
\lambda = 2^{1/3} \simeq 1.26, \qquad 
\overline E_0 = \dfrac{\sqrt{3}}{2^{4/3}} \simeq 0.69.
\end{equation}
However,  Equation \eqref{lambda-nH} will not work here because the numerator of the correction would be zero. By carefully re-doing the expansion in this special case, we find the first correction to this criterion in terms of the plate thickness to be of order one for the stretch (and the next term is of order three) and of order two for the voltage (and the next term is of order three). Explicitly, 
\begin{align}\label{thin-Hessian}
& \lambda = 2^{1/3} \pm \dfrac{2^{1/6}}{3}(\pi H/\mathcal L) \simeq 1.26 \pm 1.18 (H/\mathcal L), \notag \\[6pt] 
& \overline E_0 = \dfrac{\sqrt{3}}{2^{4/3}} - \dfrac{2^{1/3}}{3^{3/2}}(\pi H/\mathcal L)^2 \simeq 0.69 - 2.39 (H/\mathcal L)^2.
\end{align}
\begin{figure}[h!]
\centering
\includegraphics[width=0.95\textwidth]{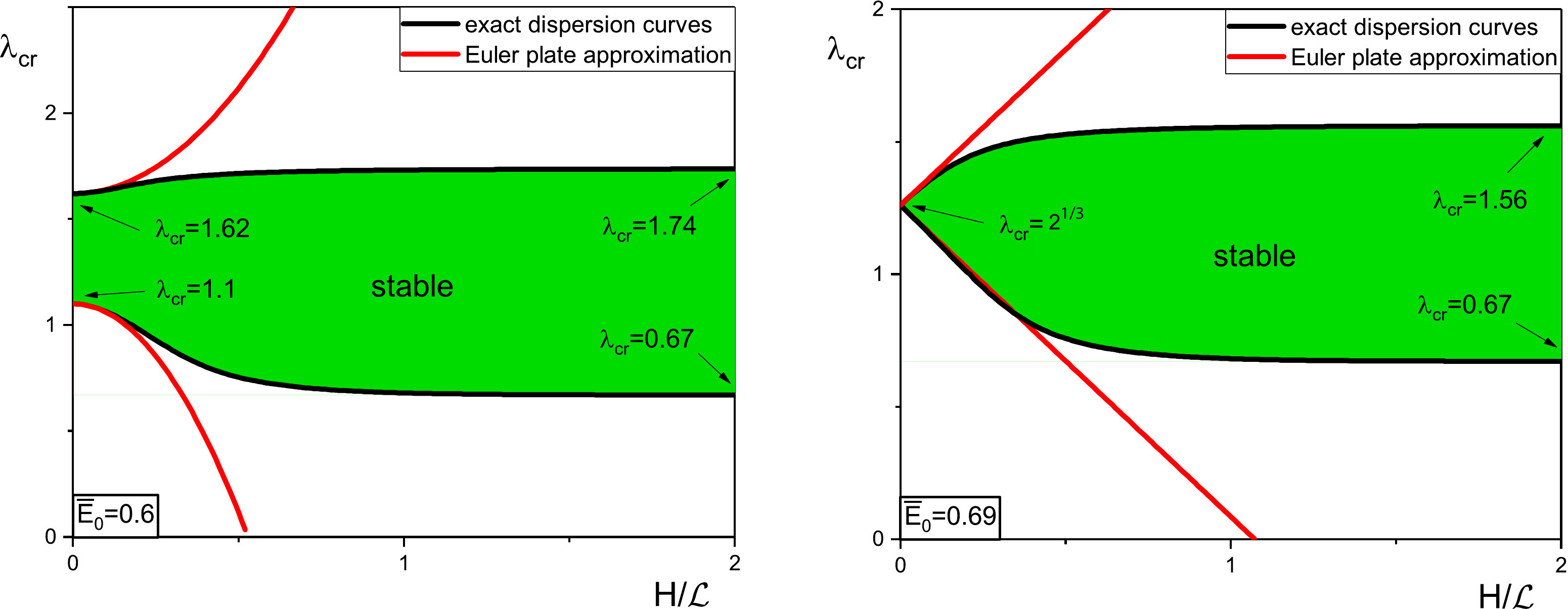}
\caption{
{\footnotesize
Critical stretch $\lambda_\text{cr}$ versus the initial thickness to wavelength ratio $H/\mathcal L$ for the anti-symmetric instability modes of a neo-Hookean ideal dielectric plate (left: $\overline E_0=0.6$, right: $\overline E_0=\sqrt{3}/2^{4/3} \simeq 0.69$).} Black: exact solutions from the analytical dispersion equation \eqref{Neo} and Red: Euler buckling approximations from Equation \eqref{lambda-nH} (left) and Equation \eqref{thin-Hessian}$_1$ (right).
}
\label{figure7}
\end{figure}

Figures \ref{figure7} show the dispersion curves obtained by the exact equation \eqref{Neo} and the Euler column buckling approximations \eqref{lambda-nH} and \eqref{thin-Hessian}$_1$ of a neo-Hookean plate. 
When $\overline E_0<0.69$ (left figure), Equation \eqref{MR-euler} has two real roots, and the $\lambda_\text{cr}-H/\mathcal L$ curves for the thin plate are approximated quadratically. 
For the  case of the Hessian instability criterion ($\overline E_0= \sqrt{3}/2^{4/3}$), Equation \eqref{MR-euler} has a single real root, and the $\lambda_\text{cr}-H/\mathcal L$ curves are approximated linearly for small thicknesses.


\section*{Acknowledgments}


This work was supported by  a visiting student scholarship for International Collaborative Research from Zhejiang University, a Government of Ireland Postgraduate Scholarship from the Irish Research Council, a Government of Ireland Postdoctoral  Fellowship from the Irish Research Council, and the National Natural Science Foundation of China (No. 11621062).




\renewcommand\thesection{\Alph{section}}
\setcounter{section}{0}
\renewcommand*{\thesection}{A}


\section{Appendix}


Here we give the general expressions for the electro-acoustic moduli and the derivation of the Stroh matrix.
The appendix is self-contained and some equations from the main text are repeated.


\subsection{Electro-acoustic moduli}


We use the push-forward versions of the incremental constitutive equations when the free energy is of the form $\Omega=\Omega({\vec {F}},\vec E_L)$ \citep{Ogden2014}: 
\begin{equation}
{\vec{ \dot T}} = \vec{\fa}_0\vec{L}+\vec{\Gamma}_0\vec{ \dot E}_L +p\vec{L}-\dot{p}\vec{I}, \qquad \DLd = - \vec{\Gamma}_0 ^T \vec{L} - \vec{K}_0\ELd,\label{constitutive}
\end{equation}
where  ${\vec{\dot T}}$ is the push-forward version of the incremental mechanical traction, $\ELd$ and $\DLd$ are the push-forward versions of the incremental electric field and electric displacement, respectively, $\vec{L} = \text{grad }\vec u$ is the gradient of the small-amplitude mechanical displacement $\vec{u}$, $p$ is the Lagrange multiplier  due to the incompressibility condition, with $\dot p$ its increment, and $\vec{\fa}_0$, $\vec{\Gamma}_0$ and $\vec{K}_0$ are fourth-, third- and second-order tensors, respectively, the so-called \emph{electro-acoustic moduli}.

The electro-acoustic moduli tensors are given in terms of the first and second derivatives of the energy density function $\Omega$ with respect to the deformation gradient $\vec{F}$ and the Lagrangian form of the electric field $\vec{E}_L=\vec{F}^T\vec{E}$, where $\vec E$ is the electric field. Using the invariants given in the main body of the paper, we obtain the following expressions for the components of the moduli tensors,
\begin{align}
\fa_{0jilk} =&  4 \left\lbrace \Omega _{11}b_{ij}b_{kl} +\Omega_{22}(I_1\vec{b}-\vec{b}^2)_{ij}(I_1\vec{b}-\vec{b}^2)_{kl}  \right. \notag \\[2.5pt]
& + \Omega _{12} \left[b_{ij}(I_1\vec{b}-\vec{b}^2)_{kl}+b_{kl}(I_1\vec{b}-\vec{b}^2)_{ij}\right] - \Omega_{15}(b_{ij}E_kE_l+b_{kl}E_iE_j) \notag \\[2.5pt]
& - \Omega_{16}\left[b_{ij}(E_k(\vec{b}^{-1}\vec{E})_l +(\vec{b}^{-1}\vec{E})_kE_l) +b_{kl}(E_i(\vec{b}^{-1}\vec{E})_j +(\vec{b}^{-1}\vec{E})_iE_j)\right] \notag \\[2.5pt]
& - \Omega_{25}\left[E_iE_j(I_1\vec{b}-\vec{b}^2)_{kl} +E_kE_l(I_1\vec{b}-\vec{b}^2)_{ij}\right] - \Omega_{26} \left[(I_1\vec{b}-\vec{b}^2)_{ij}(E_k(\vec{b}^{-1}\vec{E})_l \right. \notag \\[2.5pt]
& \left. +(\vec{b}^{-1}\vec{E})_kE_l) +(I_1\vec{b}-\vec{b}^2)_{kl}(E_i(\vec{b}^{-1}\vec{E})_j+(\vec{b}^{-1}\vec{E})_iE_j)\right] + \Omega_{55} E_iE_jE_kE_l \notag \\[2.5pt]
& + \Omega_{56} \left[E_iE_j(E_k(\vec{b}^{-1}\vec{E})_l + (\vec{b}^{-1}\vec{E})_kE_l) +E_kE_l(E_i(\vec{b}^{-1}\vec{E})_j + (\vec{b}^{-1}\vec{E})_iE_j)\right] \notag \\[2.5pt]
& \left. + \Omega_{66} (E_i(\vec{b}^{-1}\vec{E})_j + (\vec{b}^{-1}\vec{E})_iE_j)(E_k(\vec{b}^{-1}\vec{E})_l + (\vec{b}^{-1}\vec{E})_kE_l) \right\rbrace \notag \\[2.5pt]
& +2 \left\lbrace \Omega_1 \delta_{ik}b_{jl} + \Omega_2 \left( 2b_{ij}b_{kl} - b_{il}b_{jk} -b_{ik}b_{jl} +\delta_{ik}(I_1\vec{b}-\vec{b}^2)_{jl} \right) \right. \notag \\[2.5pt]
& +\Omega_5 (\delta_{jl}E_iE_k + \delta_{jk}E_iE_l + \delta_{il}E_jE_k) \notag \\[2.5pt]
& + \Omega_6 \left[b_{ik}^{-1}E_jE_l + b_{il}^{-1}E_j E_k +b_{jk}^{-1}E_i E_l + b_{jl}^{-1}E_i E_k + \delta_{jk}(E_i (\vec{b}^{-1}\vec{E})_l +(\vec{b}^{-1}\vec{E})_i E_l) \right. \notag \\[2.5pt]
& \left. \left.  + \delta_{il}(E_j (\vec{b}^{-1}\vec{E})_k + (\vec{b}^{-1}\vec{E})_j E_k) +\delta_{jl}(E_i (\vec{b}^{-1}\vec{E})_k + (\vec{b}^{-1}\vec{E})_i E_k)  \right] \right\rbrace, \label{A0jilk}
\end{align}
\begin{align}
\Gamma _{0jik} =& 4 \left\lbrace b_{ij} \left[ \Omega_{14}(\vec{bE})_k + \Omega _{15} E_k + \Omega_{16} (\vec{b}^{-1}\vec{E})_k\right] \right. \notag \\[2.5pt]
& +(I_1\vec{b}-\vec{b}^2)_{ij} \left[ \Omega_{24}(\vec{bE})_k +\Omega_{25}E_k +\Omega_{26}(\vec{b}^{-1}\vec{E})_k\right] \notag \\[2.5pt]
& -E_iE_j \left[\Omega_{45}(\vec{bE})_k +\Omega_{55}E_k +\Omega_{56} (\vec{b}^{-1}\vec{E})_k \right] \notag \\[2.5pt]
& \left. - \left( E_i(\vec{b}^{-1}\vec{E})_j + (\vec{b}^{-1}\vec{E})_iE_j \right) \left[\Omega_{46}(\vec{bE})_k +\Omega_{56}E_k +\Omega_{66} (\vec{b}^{-1}\vec{E})_k \right] \right\rbrace \notag \\[2.5pt]
& -2 \left[ \Omega_5(\delta_{jk}E_i + \delta_{ik}E_j) +\Omega_6 (\delta_{ik}(\vec{b}^{-1}\vec{E})_j +\delta_{jk}(\vec{b}^{-1}\vec{E})_i  + b^{-1} _{ik} E_j +b^{-1} _{jk} E_i) \right],\label{Gamma0jik}
\end{align}
\begin{align}
K_{0ij} =& 4 \left\lbrace \Omega_{44}(\vec{bE})_i(\vec{bE})_j +\Omega_{55} E_iE_j +\Omega_{66}(\vec{b}^{-1}\vec{E})_i(\vec{b}^{-1}\vec{E})_j \right. \notag \\[2.5pt]
& + \Omega_{45} \left[(\vec{bE})_iE_j +E_i(\vec{bE})_j \right] +\Omega_{46} \left[(\vec{bE})_i(\vec{b}^{-1}\vec{E})_j +(\vec{b}^{-1}\vec{E})_i(\vec{bE})_j \right] \notag \\[2.5pt]
& \left. + \Omega_{56} \left[ E_i(\vec{b}^{-1}\vec{E})_j +(\vec{b}^{-1}\vec{E})_iE_j \right] \right\rbrace +2 \left( \Omega_4 b_{ij} +\Omega_5 \delta_{ij} +\Omega_6 b^{-1} _{ij} \right),\label{K0ij}
\end{align}
where $\vec{b}=\vec{FF}^T$ is the left Cauchy-Green deformation tensor, $\Omega_{ij} = \partial ^2 \Omega / \partial I_i \partial I_j$ and $\Omega _i = \partial \Omega / \partial I_i$, for $i,j = 1,2,4,5,6$.

These expressions are derived from the derivatives of the invariants with respect to the deformation gradient $\vec{F}$ and to the Lagrangian electric field $\vec{E}_L$. The non-zero first derivatives are as follows, \begin{align*} \hspace{-2.5cm}
\frac{\partial I_1}{\partial F_{i\alpha}} &= 2F_{i\alpha}, 
& \frac{\partial I_2}{\partial F_{i\alpha}} &= 2(I_1 F_{i\alpha} - c_{\alpha\gamma}F_{i\gamma}), 
\end{align*}
\vspace{-0.5cm}
\begin{align*}
\frac{\partial I_5}{\partial F_{i\alpha}} &= -2c_{\alpha\gamma}^{-1}E_{L\gamma}F_{\delta i}^{-1}E_{L\delta}, &
\frac{\partial I_6}{\partial F_{i\alpha}} &= -2 \left (c_{\alpha\gamma}^{-2}E_{L\gamma}F_{\delta i}^{-1}E_{L\delta} + c_{\alpha\gamma}^{-1}E_{L\gamma}c_{\delta p}^{-1}F_{pi}^{-1}E_{L\delta} \right),
\end{align*}
\vspace{-0.5cm}
\begin{align}  \hspace{-1.3cm}
\frac{\partial I_4}{\partial E_{L\alpha}} &= 2E_{L\alpha}, 
& \frac{\partial I_5}{\partial E_{L\alpha}} &= 2c_{\alpha\gamma}^{-1}E_{L\gamma}, 
& \frac{\partial I_6}{\partial E_{L\alpha}} &= 2c_{\alpha\gamma}^{-2}E_{L\gamma},
\end{align}
the non-zero second derivatives with respect to $\vec{F}$ are
\begin{align}
\frac{\partial ^2 I_1}{\partial F_{i\alpha}\partial F_{k\beta}} =& 2\delta_{ik}\delta_{\alpha\beta}, \notag \\[4pt]
\frac{\partial ^2 I_2}{\partial F_{i\alpha}\partial F_{k\beta}} =& 2 \left( 2F_{i\alpha}F_{k\beta}-F_{i\beta}F_{k\alpha}+\delta_{ik}(I_1\delta_{\alpha\beta}-c_{\alpha\beta})-b_{ik}\delta_{\alpha\beta} \right), \notag \\[4pt]
\frac{\partial ^2 I_5}{\partial F_{i\alpha}\partial F_{k\beta}} =& 2E_{L\gamma}E_{L\delta} \left( c_{\alpha\beta}^{-1}F_{\gamma k}^{-1}F_{\delta i}^{-1} + c_{\beta\gamma}^{-1}F_{\alpha k}^{-1}F_{\delta i}^{-1} + c_{\alpha\gamma}^{-1}F_{\delta k}^{-1}F_{\beta i}^{-1} \right),\notag \\[4pt]
\frac{\partial ^2 I_6}{\partial F_{i\alpha}\partial F_{k\beta}} =& 2 \left[ c_{\alpha\beta}^{-2}F_{\gamma k}^{-1}F_{\delta i}^{-1} + c_{\beta \gamma}^{-2} F_{\alpha k} ^{-1} F_{\delta i}^{-1} + c_{\alpha\gamma}^{-2}F_{\delta k}^{-1}F_{\beta i}^{-1} + c_{\alpha \beta}^{-1} \left( c_{\gamma q}^{-1}F_{qk}^{-1}F_{\delta i}^{-1} \right. \right. \notag \\
& \left. + c_{\delta q}^{-1}F_{qi}^{-1} F_{\gamma k}^{-1} \right) + c_{\beta \gamma}^{-1} \left( c_{\alpha q}^{-1}F_{qk}^{-1}F_{\delta i}^{-1} +c_{\delta q}^{-1}F_{qi}^{-1}F_{\alpha k}^{-1} \right) \notag \\[4pt]
& \left. + c_{\alpha \gamma}^{-1} \left( c_{\delta q}^{-1}F_{qk}^{-1}F_{\beta i}^{-1} + c_{\beta q}^{-1}F_{qi}^{-1}F_{\delta k}^{-1} +c_{\beta \delta}^{-1}b_{ik}^{-1} \right) \right] E_{L\gamma}E_{L\delta}, \label{invariants2F}
\end{align}
the non-zero second derivatives with respect to $\vec{E}_L$ are 
\begin{align}
\frac{\partial ^2 I_4}{\partial E_{L\alpha}\partial E_{L\beta}} &= 2\delta_{\alpha\beta}, 
& \frac{\partial ^2 I_5}{\partial E_{L\alpha}\partial E_{L\beta}} &= 2c_{\alpha\beta}^{-1}, 
& \frac{\partial ^2 I_6}{\partial E_{L\alpha}\partial E_{L\beta}} &= 2c_{\alpha\beta}^{-2},
\end{align}
and the mixed second derivatives are 
\begin{align}
\frac{\partial ^2 I_5}{\partial F_{i\alpha}\partial E_{L\beta}} &= -2 \left( c_{\alpha\beta}^{-1}F_{\gamma i}^{-1}+c_{\alpha\gamma}^{-1}F_{\beta i}^{-1} \right) E_{L\gamma}, \notag \\[4pt]
\frac{\partial ^2 I_6}{\partial F_{i\alpha}\partial E_{L\beta}} &= -2 \left[ c_{\alpha\gamma}^{-2}F_{\beta i}^{-1} + c_{\alpha\beta}^{-2}F_{\gamma i}^{-1} + F_{pi}^{-1} \left( c_{\alpha\gamma}^{-1}c_{\beta p}^{-1} + c_{\alpha\beta}^{-1}c_{\gamma p}^{-1} \right) \right]E_{L\gamma}.
\end{align}
Note that some of these derivatives were first derived by \cite{Rudykh14}.


\subsection{Two-dimensional wrinkles in transverse electric field}


We look for two-dimensional solutions to the incremental equations so that $\vec u = \vec u(x_1, x_2)$ only.
Then $\dot{p}$, $\DLd$ and $\ELd$ are also functions of $x_1$, $x_2$ only.
Although we do not show it here, we find that this leads to $u_3=0$, $\dot{D}_{L3}=0$ and $\dot{E}_{L3}=0$ for our problem of principal wrinkles in a transverse electrical field.
Because $\curl \ELd = \mathbf 0$, we can introduce the electric potential $\varphi$ and write that
\begin{equation}
\dot{E}_{L1} = - \varphi _{,1}, \qquad \dot{E}_{L2} = - \varphi _{,2}.
\end{equation}

The push-forward versions of the incremental constitutive equations then have the following non-zero entries, \begin{align}
\dot{T}_{11} &= (\fa _{01111} +p) u_{1,1} + \fa _{01122} u_{2,2} - \Gamma _{0112} \varphi _{,2} -\dot{p}, \notag \\
\dot{T}_{12} &= (\fa _{01221} +p)u_{1,2} + \fa _{01212} u_{2,1} -\Gamma _{0211} \varphi _{,1}, \notag \\
\dot{T}_{21} &= (\fa _{01221} +p)u_{2,1} + \fa _{02121} u_{1,2} - \Gamma _{0211} \varphi _{,1}, \notag \\
\dot{T}_{22} &= (\fa _{02222}+p)u_{2,2} + \fa _{01122} u_{1,1} - \Gamma _{0222} \varphi _{,2} - \dot{p}, \label{T0}
\end{align}
and
\begin{align}
\dot{D}_{L1} &= -\Gamma _{0211} (u_{1,2} +u_{2,1}) + K _{011} \varphi _{,1}, \notag \\
\dot{D}_{L2} &= -\Gamma _{0112} u_{1,1} - \Gamma _{0222} u_{2,2} + K_{022} \varphi _{,2}, \label{DL}
\end{align}
because all other components of the electro-elastic moduli are zero. 
Here $\fa_{0jilk}, \Gamma_{0jik}$ and $K_{0ij}$ are given by Equations \eqref{A0jilk}, \eqref{Gamma0jik} and \eqref{K0ij} respectively.

The equilibrium equations in the incremental case, $\div {\vec{\dot T}} = \mathbf 0$ and $\div \DLd =  0$, are then as follows, 
\begin{align}
\dot{T}_{11,1} + \dot{T}_{21,2} = 0, \qquad
\dot{T}_{12,1} + \dot{T}_{22,2} = 0, \qquad
\dot{D}_{L1,1} + \dot{D}_{L2,2} = 0, \label{eqlb}
\end{align}
which together with the incompressibility condition, 
\begin{equation} 
\div \vec{u} = u_{1,1} + u_{2,2} = 0,\label{incomp}
\end{equation} 
fully describe the incremental motion. 


\subsection{Stroh formulation}


We look for solutions that are harmonic in the $x_1$-direction, i.e., solutions of the form \begin{equation}
\left\lbrace u_1, u_2, \dot{D}_{L2}, \dot{T}_{21}, \dot{T}_{22}, \varphi \right\rbrace = \Re \lbrace \left[k^{-1}U_1,k^{-1}U_2, \text i\Delta ,\text  i\Sigma _{21} , \text i\Sigma _{22} , k^{-1}\Phi \right]e^{\text ikx_1} \rbrace,
\end{equation}
where $U_1$, $U_2$, $\Delta$, $\Sigma_{21}$, $\Sigma_{22}$ and $\Phi$ are functions of $kx_2$ only, and $k=2\pi / \mathcal{L}$ is the wavenumber. 
We can then rewrite the full problem in Stroh form, i.e., as 
\begin{equation}
\vec{\eta} ' = \text i \vec{N}\vec{\eta}, \label{Stroh}
\end{equation}
where \begin{equation}
\vec{\eta} = \left[\begin{array}{cccccc} U_1 & U_2 & \Delta & \Sigma _{21} & \Sigma _{22} & \Phi \end{array} \right]^T = \left[ \begin{array}{cc} \vec{U} & \vec{S} \end{array} \right]^T,
\end{equation} is the Stroh vector, the prime denotes differentiation with respect to $kx_2$, and $\vec{N}$ is the Stroh matrix, which can be partitioned as  \begin{equation}
\vec{N} = \left[ \begin{array}{cc} \vec{N}_1 & \vec{N}_2 \\
\vec{N}_3 & \vec{N}_1 ^{\dagger} \end{array} \right],
\end{equation}
where $\dagger$ denotes the Hermitian operator.
We derived the Stroh matrix $\vec{N}$ as follows.

First, substituting $u_1$ and $u_2$ into the incompressibility condition, \eqref{incomp}, gives \begin{equation}
U_2 ' = -\text i U_1, \label{U2}
\end{equation}
the second line of the Stroh equation. We then substitute the expression for $\dot{D}_{L2}$ into equation \eqref{DL}$_2$ and using \eqref{U2} we get the following expression for $\Phi '$, \begin{equation}
\Phi ' = \text i \left[ \frac{\Gamma _{0112}-\Gamma_{0222}}{K_{022}}U_1 + \frac{1}{K_{022}} \Delta \right],\label{Phi}
\end{equation}
i.e., the last line of the Stroh equation.
Similarly, we can then get an expression for $U_1'$ by using $\dot{T}_{21}$ in Equation \eqref{T0}$_3$, so that 
\begin{equation}
U_1' =\text  i \left[ \frac{-(\fa_{01221} +p)}{\fa_{02121}}U_2 + \frac{1}{\fa_{02121}} \Sigma _{21} + \frac{\Gamma_{0211}}{\fa_{02121}}\Phi \right], \label{U1}
\end{equation}
which is the first line of the Stroh equation.

In order to get the remaining three equations, we  use the equilibrium equations \eqref{eqlb}. We  first find an expression for $\dot{p}$ by rearranging the expression \eqref{T0}$_4$ for $\dot{T}_{22}$, and then substitute this into \eqref{T0}$_1$ and use \eqref{Phi} and \eqref{eqlb}$_1$ to find the fourth line of the Stroh equation as follows,
\begin{multline}
\Sigma _{21} ' = -\text i \left\{ \left[ \fa_{01111} + \fa_{02222} -2\fa_{01122} +2p -\frac{(\Gamma_{0112}-\Gamma_{0222})^2}{K_{022}} \right] U_1 \right.
\\
 \left. + \Sigma_{22} -\frac{(\Gamma_{0112}-\Gamma_{0222})}{K_{022}} \Delta \right\}. \label{Sig21}
\end{multline} 
Similarly, we use \eqref{T0}$_2$, \eqref{U1} and \eqref{eqlb}$_2$ to find the fifth Stroh equation as
\begin{multline}
\Sigma_{22}' = \text i \left\{ \left[ \frac{(\fa_{01221}+p)^2}{\fa_{02121}}  -\fa_{01212}\right]U_2 - \frac{(\fa_{01221}+p)}{\fa_{02121}}\Sigma_{21} \right. \\
\left.  -\Gamma_{0211}\left( \frac{\fa_{01221}+p}{\fa_{02121}}-1\right) \Phi \right\}. \label{Sig22}
\end{multline}
Finally, to get an equation for $\Delta'$, we use \eqref{DL}$_1$, \eqref{U1} and \eqref{eqlb}$_3$ so that, 
\begin{equation}
\Delta' = \text i \left\lbrace \Gamma_{0211} \left[ 1- \frac{(\fa_{01221}+p)}{\fa_{02121}}\right]U_2 + \frac{\Gamma_{0211}}{\fa_{02121}}\Sigma_{21} + \left( \frac{(\Gamma_{0211})^2}{\fa_{02121}}-K_{011} \right) \Phi \right\rbrace .\label{Delta}
\end{equation}

We can then write these six equations in the Stroh matrix form.  
Adopting the following shorthand  notation, 
\begin{align}
& a = \fa _{01212}, && c = \fa _{02121}, && 2b = \fa _{01111} + \fa _{02222} -2\fa _{01122} -2\fa _{01221}, \notag \\
& d = \Gamma _{0211}, && e = \Gamma _{0222} - \Gamma _{0112},  && f = K_{011}, \qquad g = K_{022},\label{Stroh_const}
\end{align}
we find the partitions of the Stroh matrix, $\vec{N}_1$, $\vec{N}_2$ and $\vec{N}_3$, as follows,
\begin{align}
& \vec{N}_1 = \left[ \begin{array}{ccc}
0 & -1+\tau_{22}/c & 0 \\
-1 & 0 & 0 \\
0 & d\tau _{22} /c & 0
\end{array} \right],
\qquad
\vec{N}_2 = \left[ \begin{array}{ccc}
1/c & 0 & d/c \\
0 & 0 & 0 \\
d/c & 0 & d^2/c -f
\end{array} \right],
\notag \\[12pt]
&
\vec{N}_3 = \left[ \begin{array}{ccc}
-2(b+c-\tau_{22})+e^2/g & 0 & -e/g \\
0 & -a + (c-\tau _{22})^2/c &0 \\
-e/g & 0 & 1/g 
\end{array} \right], \label{Stroh_partitions}
\end{align}
where we have also made use of the connection $\fa _{01221} +p = \fa _{02121} - \tau _{22}$
(see \cite{Chad97} or \cite{ShDO11}). 

In particular, in this paper, we solve a problem where there is no electric field external to the plate, and so $\tau_{22}=0$ in the expressions above.

In general, the expressions \eqref{Stroh_const} read as follows,
\begin{align}
a =& 2\left[\lambda_1^2 \Omega_1 + \lambda_1 ^{2}\lambda_3 ^{2} \Omega_2 +\lambda_1^2 \lambda_3^2 E_0^2\left(\Omega_5 + (\lambda_1^{-2} + 2\lambda_1 ^{2}\lambda_3 ^{2})\Omega_6 \right) \right], \notag \\[4pt]
2b =& 4\left\lbrace \left(\lambda_1^2 - \lambda_1 ^{-2}\lambda_3 ^{-2} \right) \left[\left(\lambda_1^2 - \lambda_1 ^{-2}\lambda_3 ^{-2} \right) \left(\Omega_{11} + 2\lambda_3^{2} \Omega_{12} + \lambda_3^{4} \Omega_{22} \right) + 2\lambda_1^2 \lambda_3^2 E_0^2\left( \Omega_{15} \right. \right. \right. \notag \\
& \left. \left. \left. + 2\lambda_1 ^{2}\lambda_3 ^{2} \Omega_{16} + \lambda_3^{2} \Omega_{25} + 2\lambda_1^{2} \lambda_3^{4} \Omega_{26} \right) \right]  +\lambda_1^4 \lambda_3^4 E_0^4 \left( \Omega_{55} + 4\lambda_1 ^{2}\lambda_3 ^{2} \Omega_{56} + 4\lambda_1 ^{4}\lambda_3 ^{4} \Omega_{66} \right) \right\rbrace \notag \\
& +2\left\lbrace \left(\lambda_1^2 + \lambda_1 ^{-2}\lambda_3 ^{-2} \right) \left(\Omega_1 + \lambda_3^{2}  \Omega_2 \right) +\lambda_1^2 \lambda_3^2 E_0^2\left[ \Omega_5 + 2(3\lambda_1 ^{2}\lambda_3 ^{2} - \lambda_1^{-2})\Omega_6 \right] \right\rbrace , \notag \\[4pt]
c =& 2\left[ \lambda_1 ^{-2}\lambda_3 ^{-2} \Omega_1 + \lambda_1^{-2} \Omega_2 + \lambda_3^{2} E_0 ^2 \Omega_6 \right], \notag \\[4pt]
d =& -2\lambda_1 \lambda_3 \left[ \Omega_5 + (\lambda_1^{-2} + \lambda_1 ^{2}\lambda_3 ^{2}) \Omega_6 \right] E_0, \notag \\[4pt]
e =& 4 \lambda_1 \lambda_3 \left[ (\lambda_1 ^{-2}\lambda_3 ^{-2} - \lambda_1^2)(\lambda_1 ^{-2}\lambda_3 ^{-2} \Omega_{14} + \Omega_{15} + \lambda_1 ^{2}\lambda_3 ^{2} \Omega_{16} + \lambda_1^{-2} \Omega_{24} + \lambda_3 ^{2} \Omega_{25} + \lambda_1^{2} \lambda_3^{4} \Omega_{26}) \right. \notag \\
& \left. - \lambda_1 ^2 \lambda_3^2 E_0^2 (\lambda_1 ^{-2}\lambda_3 ^{-2}\Omega_{45} +2\Omega_{46} + \Omega_{55} +3\lambda_1 ^{2}\lambda_3 ^{2}\Omega_{56} +2\lambda_1 ^{4}\lambda_3 ^{4} \Omega_{66}) -(\Omega_5 + 2\lambda_1 ^{2}\lambda_3 ^{2} \Omega_6) \right] E_0, \notag \\[4pt]
f =& 2(\lambda_1^2 \Omega_4 + \Omega_5 + \lambda_1 ^{-2} \Omega_6), \notag \\[4pt]
g =& 4 \left[ \lambda_1 ^{-4}\lambda_3 ^{-4} \Omega_{44} + 2\lambda_1 ^{-2}\lambda_3 ^{-2} \Omega_{45} + 2\Omega_{46} +\Omega_{55} + 2\lambda_1 ^{2}\lambda_3 ^{2} \Omega_{56} + \lambda_1 ^{4}\lambda_3 ^{4} \Omega_{66} \right] \lambda_1^2 \lambda_3^2 E_0 ^2 \notag \\
& +2(\lambda_1 ^{-2}\lambda_3 ^{-2} \Omega_4 + \Omega_5 + \lambda_1 ^{2}\lambda_3 ^{2} \Omega_6).
 \label{gen_const}
\end{align}

We can \emph{non-dimensionalise} both the Stroh constants and entries of $\vec{\eta}$ by introducing the following dimensionless moduli,
\begin{align}
&\overline{a} = a/\mu, && \overline{b} = b/\mu, && \overline{c} = c/\mu, && \overline \tau_{22} = \tau_{22}/\mu, \notag \\
& \overline{d} = d/\sqrt{\mu\varepsilon}, && \overline{e} = e/\sqrt{\mu\varepsilon},
&&\overline{f} = f/\varepsilon, && \overline{g} =g/\varepsilon, \label{dimls_const}
\end{align}
and dimensionless fields,
\begin{equation}
\overline{U}_i = U_i, \qquad \overline{\Sigma}_{2i} = \Sigma_{2i}/\mu, 
\qquad 
\overline{\Delta} = \Delta/\sqrt{\mu\varepsilon},
\qquad
\overline{\Phi} = \Phi\sqrt{\varepsilon/\mu}, 
\label{dimls_eta}
\end{equation}
for $i=1,2$, where $\overline{X}$ denotes a dimensionless measure of $X$, and $\mu$ and $\varepsilon$ are the initial shear modulus and initial permittivity of the dielectric material,
\begin{equation}
\mu = 2(\Omega_1+\Omega_2)|_{I_1=I_2=3,  I_4=I_5=I_6=0}, \qquad
\varepsilon = -2(\Omega_4+\Omega_5+\Omega_6)|_{I_1=I_2=3,  I_4=I_5=I_6=0}.
\end{equation}
The finite fields can also be non-dimensionalized by introducing
\begin{equation}
\overline E_0 = E_0\sqrt{\varepsilon/\mu}, \qquad
\overline D = D/\sqrt{\mu \varepsilon}, \qquad
\overline I_\alpha = (\varepsilon/\mu) I_\alpha  \quad (\alpha=4,5,6).
\end{equation}

Once $a$ is replaced by $\mu \overline a$, $b$ by $\mu \overline b$, etc., the equations of equilibrium can be re-written in their non-dimensional form $\vec{\overline \eta}' = \text i \vec{\overline N} \vec{\overline \eta}$. 
For the rest of the appendix, the overline notation is understood everywhere, and \emph{all quantities are non-dimensional}.


\subsection{Method of resolution for plates}


Since $\vec{N}$ has constant entries, we look for solutions to \eqref{Stroh} in the form, 
\begin{equation}
\vec{\eta}(kx_2) = \vec{\eta}^0 e^{\text iqkx_2},
 \label{eta}
\end{equation} 
which results in an eigen-problem for the eigenvalues $q$ and eigenvectors $\vec{\eta}^0$ of the matrix $\vec{N}$, \begin{equation}
(\vec{N}-q\vec{I})\vec{\eta}^0=\vec{0}.
\end{equation} 
The characteristic equation associated with this eigen-problem is
\begin{equation}
cgq^6 + [2bg+cf-(d-e)^2]q^4 + [2bf +ag+2d(d-e)] q^2+af-d^2=0. \label{ch-eqn}
\end{equation}
This equation is bi-cubic in $q$ and does not depend on the Cauchy stress $\tau _{22}$ for any choice of energy density function.

After calculating the eigenvalues $q_j$ and eigenvectors $\vec{\eta}^{(j)}$, $j=1,2, \dots, 6$, for the Stroh matrix $\vec{N}$, we can construct the solution to \eqref{Stroh} for a plate of electroelastic material, \begin{equation}
\vec{\eta}(kx_2) = \left[ \begin{array}{c} \vec{U}(kx_2) \\
\vec{S}(kx_2) \end{array} \right] = \sum _{j=1} ^6 c_j \vec{\eta} ^{(j)} e^{\text iq_jkx_2}, \label{eta_plate}
\end{equation}
where $c_j$ for $j=1,2,\dots,6$ are arbitrary constants to be determined from the boundary conditions. 
The eigenvalues come in conjugate pairs because the bicubic has real coefficients. 
We specialise the analysis to free energies for which the $q_j$ are pure imaginary, and so we write them as $q_j = \text ip_j$ and $q_{j+3} = - \text ip_j$ for $j=1,2,3$,  where $p_1$, $p_2$, $p_3$ are real. 
Then the eigenvectors are also conjugate pairs, $\vec{\eta}^{(j)} = \overline{\vec{\eta}^{(j+3)}}$ for $j=1,2,3$.  

The incremental equations must be solved subject to the boundary conditions of no incremental mechanical tractions and no incremental electric field on the faces of the plate, i.e., $\vec{S}(kh/2) = \vec{S}(-kh/2) = \vec{0}$. 
Using this boundary condition and \eqref{eta_plate}, we can write the following matrix equation, \begin{equation} \left[ \begin{array}{c} \vec{S}(kh/2) \\
\vec{S}(-kh/2) \end{array} \right] = \left[ \begin{array}{cccccc} F_1E_1 ^- & F_2E_2 ^- & F_3E_3 ^- & F_4E_1 ^+ & F_5E_2 ^+ & F_6E_3 ^+ \\
G_1E_1 ^- & G_2E_2 ^- & G_3E_3 ^- & G_4E_1 ^+ & G_5E_2 ^+ & G_6E_3 ^+ \\
H_1E_1 ^- & H_2E_2 ^- & H_3E_3 ^- & H_4E_1 ^+ & H_5E_2 ^+ & H_6E_3 ^+ \\
F_1E_1 ^+ & F_2E_2 ^+ & F_3E_3 ^+ & F_4E_1 ^- & F_5E_2 ^- & F_6E_3 ^- \\
G_1E_1 ^+ & G_2E_2 ^+ & G_3E_3 ^+ & G_4E_1 ^- & G_5E_2 ^- & G_6E_3 ^- \\
H_1E_1 ^+ & H_2E_2 ^+ & H_3E_3 ^+ & H_4E_1 ^- & H_5E_2 ^- & H_6E_3 ^-  \end{array} \right] \left[ \begin{array}{c} c_1 \\ c_2 \\ c_3 \\ c_4 \\ c_5 \\ c_6 \end{array} \right] = \vec{0}, \label{full_bc}
\end{equation}
where $F_j = \eta _4 ^{(j)}$, is the fourth component, $G_j = \eta _5 ^{(j)}$ is the fifth component, and $H_j = \eta _6 ^{(j)}$ is the sixth component of the eigenvector $\vec{\eta}^{(j)}$, and $E_j ^{\pm} = e^{\pm p_j kh/2}$, for $j=1,2,\dots,6$. We also note that since $q_{j+3} = -q_j$, we have $E_{j+3} ^{\pm} = E_j^{\mp}$, for $j=1,2,3$. 

For our choices of free energy densities, we find that $F_{j+3}=F_j$, $G_{j+3}=-G_j$, $H_{j+3}=H_j$, for $j=1,2,3$. 
Then some simple linear manipulations \citep{Nayfeh95} of the matrix result in two $3 \times 3$ blocks, the antisymmetric and symmetric modes, and its determinant factorises as follows, \begin{equation}
\left| \begin{array}{ccc} 
F_1C_1 & F_2C_2 & F_3C_3 \\
G_1S_1 & G_2S_2 & G_3S_3 \\
H_1C_1 & H_2C_2 & H_3C_3 \end{array} \right|
\times  
\left| \begin{array}{ccc} 
F_1S_1 & F_2S_2 & F_3S_3 \\
G_1C_1 & G_2C_2 & G_3C_3 \\
H_1S_1 & H_2S_2 & H_3S_3 \end{array} \right| = 0, 
\end{equation}
where $C_j = \cosh(p_jkh/2)$ and $S_j = \sinh(p_jkh/2)$.
Here the antisymmetric mode is described by the determinant on the left, and the symmetric mode by the one on the right. We then get the following expressions for the dispersion equations in general, 
\begin{multline}
G_1(F_3H_2-F_2H_3)\tanh (p_1kh/2) + G_2(F_1H_3-F_3H_1)\tanh(p_2kh/2)  \\
+ G_3(F_2H_1-F_1H_2)\tanh(p_3kh/2)=0, \label{antisym_disp}
\end{multline} for the antisymmetric mode and, 
\begin{multline}
G_1(F_3H_2-F_2H_3)\coth (p_1kh/2) + G_2(F_1H_3-F_3H_1)\coth(p_2kh/2)  \\
+ G_3(F_2H_1-F_1H_2)\coth(p_3kh/2)=0, \label{sym_disp}
\end{multline} for the symmetric mode.
The quantity $kh$ can be expressed as $kh = 2 \pi \lambda_1^{-1} \lambda_3^{-1} H/\mathcal L$, where $H$ is the initial thickness of the plate and $\mathcal L$ is the wavelength of the wrinkles, two quantities that are easy to measure experimentally.


\subsection{Examples}


Of course, solving the bicubic  \eqref{ch-eqn} is quite cumbersome in the general case, but for some special forms of the free energy density, it simplifies quite a lot. 
Hence, for \emph{generalized neo-Hookean ideal dielectrics}, which are such that
\begin{equation} \label{ideal}
\Omega = W(I_1) - \dfrac{\varepsilon}{2} I_5,
\end{equation}
where $W$ is an arbitrary function of $I_1$ only, we find that it factorises as
\begin{equation}
(q^2+1)\left\{q^4 + \left[1 + \lambda_1^4\lambda_3^2 + 2 (\lambda_1^3\lambda_3 - \lambda_1^{-1}\lambda_3^{-1})^2 \dfrac{W''}{W'}\right]q^2+ \lambda_1^4\lambda_3^2 \right\} = 0.
\end{equation}
Here we call $q_1$ and $q_2$ the two roots of the factorized biquadratic with positive imaginary part.

Now we define all six eigenvalues and the three real numbers $p_j$ by 
 \begin{equation}
 q_1=-q_4=\ii p_1, \qquad q_2=-q_5=\ii p_2, \qquad q_3=-q_6=\ii \quad (p_3=1).
 \end{equation}
 The real quantities $p_1$, $p_2$ are such that
\begin{equation}
p_1^2 p_2^2 = \lambda_1^4\lambda_3^2, \qquad
p_1^2 + p_2^2 = 1 + \lambda_1^4\lambda_3^2 + 2( \lambda_1^3\lambda_3 - \lambda_1^{-1}\lambda_3^{-1})^2 \dfrac{W''}{W'}.
\end{equation}
Solving for real positive $p_1$, $p_2$ gives
\begin{equation}
p_{1,2} = \dfrac{\lambda_1^2\lambda_3 + 1}{2}\sqrt{1 + 2(\lambda_1 - \lambda_1^{-1}\lambda_3^{-1})^2\dfrac{W''}{W'}} \pm  \dfrac{\lambda_1^2\lambda_3 - 1}{2}\sqrt{1 + 2(\lambda_1 + \lambda_1^{-1}\lambda_3^{-1})^2\dfrac{W''}{W'}}.
\end{equation}

The  six  eigenvectors are
\begin{align}
& \vec{\eta}^{(1)}
=\left[ \begin{matrix}
-\ii{p_1} 
\\[4pt]
1 
\\[4pt]
-\ii p_1 \lambda_1\lambda_3E_0 
\\[12pt]
 2(1+p_1^2)\dfrac{W'}{\lambda_1^2\lambda_3^2} +\lambda_1^2\lambda_3^2 E_0^2
\\[15pt]
 2\ii \dfrac{(\lambda_1^4\lambda_3^2+p_1^2)}{p_1\lambda_1^2\lambda_3^2} W' 
\\[12pt]
-\lambda_1\lambda_3 E_0
\end{matrix} \right], 
\qquad
\vec{\eta}^{(2)}
=\left[ \begin{matrix}
-\ii{p_2} 
\\[4pt]
1 
\\[4pt]
-\ii p_2 \lambda_1\lambda_3E_0 
\\[12pt]
 2(1+p_2^2)\dfrac{W'}{\lambda_1^2\lambda_3^2} +\lambda_1^2\lambda_3^2 E_0^2
\\[15pt]
 2\ii \dfrac{(\lambda_1^4\lambda_3^2+p_2^2)}{p_2\lambda_1^2\lambda_3^2}W' 
\\[12pt]
-\lambda_1\lambda_3 E_0
\end{matrix} \right], 
 \notag \\
& \vec{\eta}^{(3)}= \left[ \begin{matrix}
0 \\
0  \\
\ii\\
\lambda_1\lambda_3  E_0\\
\ii\lambda_1\lambda_3 E_0\\
-1  
\end{matrix} \right],
\qquad
\vec{\eta}^{(4)} = \overline{\vec\eta^{(1)}}, 
\qquad
\vec{\eta}^{(5)} = \overline{\vec\eta^{(2)}}, 
\qquad
\vec{\eta}^{(6)} = \overline{\vec\eta^{(3)}}.
\end{align}
From these expressions we deduce the dispersion equation for anti-symmetric buckling \eqref{antisym_disp} in the form
\begin{multline}
 2   W' \left[p_1 (1 + p_2^2)^2 \tanh\left(\pi p_1 \lambda_1^{-1}\lambda_3^{-1} H/\mathcal L \right) - p_2 (1 + p_1^2)^2\tanh\left(\pi p_2 \lambda_1^{-1}\lambda_3^{-1}H/\mathcal L \right)\right]   \\[4pt]
 = ( p_2^2 - p_1^2) \lambda_1^{4}\lambda_3^{4}  E_0^2 \tanh \left( \pi \lambda_1^{-1}\lambda_3^{-1}H / \mathcal L \right).
\end{multline}
Here we can take $H/\mathcal L \to 0$ and $H/\mathcal L \to \infty$ to establish explicit expressions for the thin-plate and the short-wave limits, as in the main text. 
The equation is valid for any  plate made of a generalized neo-Hookean ideal dielectric \eqref{ideal}, subject to a bi-axial pre-stretch $\lambda_1$, $\lambda_3$. 

For instance, for a neo-Hookean ideal dielectric plate (Equation \eqref{nH}) in the plane strain $\lambda_1=\lambda$, $\lambda_3=1$, we have $W'=1/2$, $W''=0$, $p_1=\lambda^2$, $p_2=1$, and the dispersion equation simplifies to
\begin{equation}
\dfrac{\tanh(\pi \lambda H/\mathcal L)}{\tanh(\pi \lambda^{-1}H/\mathcal L)} = 
\dfrac{(1 + \lambda^4)^2}{4\lambda^2} + \dfrac{\lambda^2 - \lambda^6}{4} E_0^2,
\end{equation}
which recovers the purely elastic buckling criterion when $E_0=0$ \citep{OgRo93}. 
However, we cannot compare it directly with the result of \cite{Yang17} when $E_0 \ne 0$, because it was obtained for different incremental electric boundary conditions.

Another example of free energy density for which it is possible to make good progress is defined by the following class,
\begin{equation}
\Omega = \frac{\mu(1-\beta)}{2}(I_1 -3) + \frac{\mu \beta}{2}(I_2-3) - F(I_5),
\end{equation}
where $F$ is an arbitrary function of $I_5$ only and $1 \ge \beta \ge 0$. 
Then we find that the characteristic equation \eqref{ch-eqn} factorises fully, as \begin{equation}
(q^2 + 1)(q^2 + \lambda_1^4 \lambda_3^2)\left[(2\lambda_1^{2}\lambda_3^2E_0^2F'' +F')q^2 + F'\right] = 0.
\end{equation}
The six eigenvalues can again be written in terms of three real numbers $p_j$,
 \begin{align}
q_1 &= -q_4 = \ii p_1, & q_2 &= -q_5 = \ii p_2, & q_3 &= -q_6 = \ii p_3,
\end{align} 
where 
\begin{equation}
p_1=1, \qquad
p_2 = \lambda_1^2\lambda_3, \qquad
p_3 = \sqrt{\dfrac{F'}{2\lambda_1^{2}\lambda_3^2 F'' E_0^2  + F'}}.
\end{equation}

We find that the corresponding eigenvectors are 
\begin{align}
& \vec{\eta}^{(1)} =
  \left[ \begin{array}{c} \ii \\[2pt]
-1 \\[2pt]
-2\ii  \lambda_1 \lambda_3 F' E_0 \\[2pt]
-2 \lambda_1^{2}\lambda_3^2 F' E_0^2  -2 \lambda_1^{-2}\kappa \\[2pt]
-\ii \lambda_3 ^2 (\lambda_1^2 + \lambda_1^{-2}\lambda_3^{-2}) \kappa \\[2pt]
\lambda_1 \lambda_3 E_0 \end{array} \right],
 \qquad 
\vec{\eta}^{(2)} = 
   \left[ \begin{array}{c} \ii \lambda_1 \\[2pt]
-\lambda_1^{-1}\lambda_3^{-1} \\[2pt]
2\ii  \lambda_1^2 \lambda_3 F' E_0  \\[2pt]
-2 \lambda_1 \lambda_3 F' E_0^2  -\lambda_1 ^{-1} \lambda_3 (\lambda_1^2 + \lambda_1^{-2}\lambda_3^{-2}) \kappa \\[2pt]
-2\ii \lambda_1^{-1}\kappa \\[2pt]
E_0 \end{array} \right], \notag \\[6pt]
&
\vec{\eta}^{(3)} =
 \left[ \begin{array}{c} 
0 \\[2pt]
0 \\[2pt]
-2\ii   F' \\[2pt]
-2 p_3 \lambda_1 \lambda_3 F' E_0 \\[2pt]
-2\ii  \lambda_1 \lambda_3 F' E_0 \\[2pt]
p_3 \end{array} \right],
\qquad
\vec{\eta}^{(4)} = \overline{\vec{\eta}^{(1)}}, 
\qquad
\vec{\eta}^{(5)} = \overline{\vec{\eta}^{(2)}}, 
\qquad
\vec{\eta}^{(6)} = \overline{\vec{\eta}^{(3)}}.
\end{align}
where 
\begin{equation} 
\kappa = \lambda_3^{-2} + \beta(1-\lambda_3^{-2}). 
\end{equation}

We then find that the antisymmetric mode for the \emph{thin-plate limit}, $a-c=0$, in this case reads
\begin{equation}
2\lambda_1^2 F' E_0^2 + \kappa (\lambda_1 ^2 - \lambda_1^{-2}\lambda_3^{-2}) = 0,
\end{equation}
and that the \emph{short-wave limit}, $H/\mathcal L \to \infty$, is  as follows,
\begin{equation}
p_3 \kappa (\lambda_1^3 + \lambda_1 \lambda_3^{-1} +3\lambda_1^{-1}\lambda_3^{-2} - \lambda_1^{-3} \lambda_3^{-3}) - 2\lambda_1^2(\lambda_1+\lambda_1^{-1}\lambda_3^{-1}) F' E_0^2 = 0. \label{short-wave-f}
\end{equation}

\end{document}